\documentclass[12pt,emtex]{article}
\usepackage{epsfig,latexsym}
\psfigdriver{dvips}
\begin{document}

\title{Equal-time hierarchies for transport descriptions of fermionic fields
}
\author{
R.A. Ionescu$^1$,  H.H. Wolter\\
\normalsize{Sektion Physik, Universit\" at M\" unchen, D-85748 Garching, 
Germany} 
}
\date{\normalsize \today}
\maketitle
\begin{abstract}

A transport theory which is not restricted to the gradient  and
quasi-particle approximations is
presented
which is formulated in terms of the energy moments, or equivalently
the equal-time derivatives of the one-particle Green functions. 
A set of infinite hierarchies of 
kinetic and constraint
equations for equal-time
quantities for the spectral and the kinetic part of the 
one-particle Green function is derived. The hierarchies for the 
spectral function truncate automatically as in the mean field 
approximation.  The possibility of a systematic truncation of the 
hierarchies for the kinetic part is discussed. The effects of the 
quantum corrections are illustrated  in a simple 
one-dimensional model.

\end{abstract}

\setlength{\baselineskip} {4ex}

\footnotetext[1]{Permanent Adress: National Institute for Physics and
Nuclear Engineering "H. Hulubei", P.O.Box
MG-6, Bucharest, Romania; e-mail: amilcar@tandem.nipne.ro}
\newpage

\section{Introduction}

A transport description of many-body non-equilibrium systems is a 
very convenient 
approximation  used with success in the study of heavy ion collisions in
the last two decades \cite{THI}. It has also been used for more than 40 years
in solid state physics 
to describe the transport properties like electric conductivity, thermal
conductivity, etc. In this approach the one-body
phase-space distribution function satisfies a semi-classical transport
equation of Boltzmann type. The validity of a transport description
can be established by a rigorous derivation from quantum field theory
\cite{QP}.

A main step in the derivation
of transport equations starting from 
the Dyson equation for the Green function
is to put particles on the mass-shell \cite{QP}. On the other hand,
off-shell
effects are thought to be important  in  heavy ion collisions for processes
like subthreshold production of particles or for the inclusion of baryonic
resonances in the transport description \cite{OFF}. This is the case,
in particular, for the
$\Delta$-resonance for which the spectral function has a large width even
outside the medium.

The off-shell effects are related to the dependence of one-particle Green 
functions on two time arguments.
A sytematic way to include the off-shell effects into the dynamics 
of a particle
in an external {\it c}-number field (or in the mean field approximation) was
developed \cite{ETH} resulting in transport {\it and} constraint equations for
equal time quantities which are derivatives of the Green function
in the relative time at zero relative time, or equivalently energy moments 
of its Wigner transform.
The main advantage of this equal-time description is the
possibility to specify the initial conditions necessary to solve the 
transport equations.

More recently new derivations of semiclassical transport equations 
\cite{CAS,LEU,KNOLL}
do not use the on-shell approximation. They result in first order
partial differential equation for some function defined on 8-dimensional
(phase) space which is numerically solved using the test particle method
by further approximating it by a sum of $\delta$ functions.
It has been argued in ref. \cite{LEU} that in general the function
that can be represented by a test particle ansatz is not the phase
space distribution function, but some more general quantity.
This complicates the solution of the transport equations in 8-dimensional
phase space, giving interest to alternative methods.
%This function   is different in different approaches \cite{LEU}. 
In addition these treatments use
the first-order gradient approximation which is based on 
the assumption of
a weakly inhomogeneous and slowly evolving system. 
The gradient approximation may loose validity
when the range of the 
interactions becomes comparable with that of the space inhomogeneities
of the system. 
Other authors \cite{JOY} therefore generalize the Boltzmann equation
by including third order terms
in the space gradients and obtain equations for quantities which
depend on position, time, energy and two momentum components.

In the present work we propose a systematic way to derive equations for
quantities which characterize the off-shell behaviour of the Green functions 
and also include higher order gradient terms, {\it i.e.} go
beyond the mean field approximation. 
The idea is to formulate a theory for the energy moments of the 
one-particle Green functions. These energy moments measure the shape of 
the Green functions in energy, {\it e.g.} the second order energy moment
is a measure of the width of the spectral function. Generally the
$n^{th}$-order energy  moment is related to  the $n^{th}$-order
derivative of the Green function with respect to the relative time 
taken at zero relative time. 
We  use the method of real-time Green functions defined on the 
Schwinger-Keldysh time path, which is the appropriate formalism to describe 
quantum systems in non-equilibrium.
Taking the Wigner transform of the Dyson equation in space coordinates only 
we obtain evolution and constraint equations for the energy moments of 
Green functions.
The result is a set of four infinite hierarchies of equations. 
A systematic way to truncate these hierarchies is discussed.
%This approach is illustrated in the
%present paper for a fermionic Schr\"{o}dinger field. 

As we are going beyond the first order gradient approximation, we can study 
the non-localization effects due to the quantum motion.
There have been attempts to include the nonlocal terms in the 
scattering integral \cite{MOR} and this leads to consistency with the 
thermodynamic virial corrections .

%Our appoach can be used to study both weakly inhomogeneous but not
%slowly evolving system (using a first-order gradient approximation in 
%space only), and not weakly inhomogeneus but slowly evolving system
%(going beyond the first-order gradient approximation). 

In the last section we will illustrate 
in a simple 1-dimensional example the effects of the higher 
order gradient terms in the evolution equation as well as the influence 
of the memory effects on the velocity of thermalization.

\vskip 1truecm
\section{Spatial Wigner transforms of Dyson equations}

To keep the formalism simple 
we consider in this work a non-relativistic system of fermions 
described by the Hamiltonian
\begin{eqnarray}
H & =&  \int d^3 x \Psi_{H} ^{\dagger} (x) \left( - {{1}\over{2m}}
\bigtriangleup \right) \Psi_{H} (x) \nonumber \\ &&
+  {{1}\over{2}} \int d^3 x \int d^3 y \Psi_{H} ^{\dagger} (x)
\Psi_{H} ^{\dagger} (y) V(x-y) \Psi_{H}(y) \Psi_{H}(x)
\end{eqnarray}
where $\Psi_{H} ^{\dagger}$ and $\Psi_{H}$ are the field operators in the 
Heisenberg picture and $V$ is the two-particle interaction potential
($\hbar=1$).
A Dirac or Klein-Gordon field can be treated along the same lines, 
however, here we
want to avoid  the technical complications related to the Dirac matrix
structure of the Green function and self-energy.
%The Dyson equation is derived by a perturbation expansion
%in the interaction picture \cite{PERT} for the time evolution operator
%\begin{eqnarray}
%U_I (t_1, t_2 ) & = & \left\{  \begin{array}{cc}
%                 T^c e^{-i \int_{t_2}^{t_1} dt H_{I}' (t)} &
%\quad\mbox{if}\quad t_1 > t_2       \\
%                 T^a e^{-i \int_{t_2}^{t_1} dt H_{I}' (t)} &
%\quad\mbox{if}\quad t_1 < t_2
%                               \end{array} \right. \hfill
%\end{eqnarray}
%where $H_I '(t)$ is the interaction Hamiltonian in the interaction picture
%and $T^c$ and $T^a$ are, respectively, the chronological and anti-chronological
%time-ordering operators.

For a non-equilibrium situation it is convenient to introduce 
a path-ordered Green
function defined on the Schwinger-Keldysh double time contour \cite{KEL}. 
The one-particle 
Green function on the contour
${G}$
is defined
as the expectation value of a product of field operators in the Heisenberg
picture
\begin{eqnarray}
i {G}(1,1') & = &  Tr \left[ \rho_H P(
\Psi_H(1) \Psi_H^{\dagger}(1')) \right] \, ,
\end{eqnarray}
where $1 \equiv ({\bf x}_1, t_1)$ and $\rho_H$ is the time-independent
density operator which specifies the state. Depending on  the
position of the two time arguments on the contour ${G}$ 
is equivalent to four different
Green functions defined on the real time axis: causal and anticausal when
the two time arguments are on the same branch of the contour and correlation
Green functions when the time arguments are  on different branches
\begin{eqnarray}
i G^{c,a}(1,1') & = &  Tr \left[ \rho_H T^{c,a}(
\Psi_H(1) \Psi_H^{\dagger}(1')) \right] \nonumber \\
i G^{>}(1,1') & = &  Tr \left[ \rho_H (
\Psi_H(1) \Psi_H^{\dagger}(1')) \right] \hfill \\
i G^{<}(1,1') & = & - Tr \left[ \rho_H (
\Psi_H ^{\dagger}(1') \Psi_H (1)) \right] \nonumber
\end{eqnarray}
The causal and anticausal Green functions satisfy the following decomposition
relations
\begin{eqnarray}
G^{c,a}(1,1') & = & \theta (t_1 - t_{1'}) G^{>,<} (1,1') +
\theta ( - (t_1 - t_{1'})) G^{<,>} (1,1')
\end{eqnarray}
It is useful to define retarded and advanced Green functions
\begin{eqnarray}
G^{+,-}(1,1') & = & \pm \theta ( \pm (t_1 - t_{1'}))
\left[ G^{>} (1,1') -G^{<} (1,1') \right]
\end{eqnarray}
as well as spectral and Keldysh (or kinetic) Green functions
\begin{eqnarray}
G^{A}(1,1') & = & G^{>} (1,1') - G^{<} (1,1')   \\
G^{K}(1,1') & = & G^{>} (1,1') + G^{<} (1,1')  \, .
\end{eqnarray}
We note \cite{QP}
that all the Green functions defined above can be 
expressed in terms of only two independent
Green functions, for which we take
$G^{A}$ and $G^{K}$.
For future reference we note the relations
\begin{eqnarray}
{( G^{K,A}(1,1') )}^{\star} & = &  - G^{K,A} (1',1)
\end{eqnarray}
and, as a result of the equal-time commutation relations, the property
\begin{eqnarray}
G^{A}(1,1') \mid _{t_1 = t_{1'}} & = &  \delta ( {\bf x}_1 -   {\bf x}_{1'} )
\, .
\end{eqnarray}
Collecting the Green functions into a $2 \times 2$ matrix
$\underline{G} = \left(
\begin{array}{cc}
G^c & G^< \\
G^> & G^a
\end{array}
\right)$,
 and correspondingly for the other one-particle quantities,
the Dyson equation can be written in the equivalent forms
\begin{eqnarray}
\underline{G} (1,1') & = &   {\underline{G}}_0 (1,1') + \int d1' \int d1''
{\underline{G}}_0 (1,1'') \tilde{\underline{\Sigma}} (1'',1''')
\underline{G} (1''',1') \\
\underline{G} (1,1') & = &   {\underline{G}}_0 (1,1') + \int d1' \int d1''
{\underline{G}} (1,1'') \tilde{\underline{\Sigma}} (1'',1''') {\underline{G}}_0 (1''',1')
\end{eqnarray}
where the time arguments take values on the Schwinger-Keldysh contour and
${\underline{G}}_0$ is the free Green function which satisfies the equations
\begin{eqnarray}
( i {{\partial}\over{\partial t_1}} + {{1}\over{2m}}
{\bigtriangleup}_{1} )  {\underline{G}}_0 (1,1') & = &
\underline{\delta} (1,1') \nonumber \\
( - i {{\partial}\over{\partial t_{1'}}} + {{1}\over{2m}}
{\bigtriangleup}_{1'} )  {\underline{G}}_0 (1,1') & = &
\underline{\delta} (1,1')
\end{eqnarray}
with $\underline{\delta}$ the Dirac $\delta$-function defined on the contour.
The self-energy $\tilde{\underline{\Sigma}}$ contains all the information 
on  the interaction. 
The way in which it is approximated specifies the type
of approximation used. We will discuss this for a few cases below.
Its causal and anti-causal components have the same decomposition 
relations as the
Green functions, eq. (5), provided the singular, {\it i.e.} time-local, 
part on the contour is separated off
\cite{DEC}
\begin{eqnarray}
\tilde{\underline{\Sigma}} (1, 1') & = & 
{\Sigma}_{\delta}({\bf x}_1 , {\bf x}_{1'},t_1)
\underline{\delta}(t_1 - t_{1'}) + \underline{\Sigma} (1, 1')
\end{eqnarray}
This singular part is given by the Hartree and Fock diagrams
which are local in time, and are also discussed further in Sect. 5 below. 
Also for $\underline{\Sigma}$ we apply  analogous definitions as in eqs. (6,7).
In the following, to avoid writing integrals, we  consider the space-time 
coordinates as matrix indices
and use the Einstein convention to sum (integrate) over the repeated indices 
(coordinates).

The equations satisfied by the interacting Green functions, eqs. (10) and (11),
can be put in  a triangular matrix form for 
$\tilde{\underline{G}}=\left(
\begin{array}{cc}
G^- & 0 \\
G^K & G^+
\end{array}
\right)$
\cite{TRI}. From the non-diagonal part we obtain the 
equations for the kinetic part of the Green function
\begin{displaymath}
s _1 G^K (1,1')
-\Sigma_\delta(1 , {1''})
%{\delta}(t_1 - t_{1''})  
 G^K (1'',1')
= \Sigma^+(1,1'') G^K (1'',1') + \Sigma^K(1,1'') G^- (1'',1')
\end{displaymath}
\begin{equation}
s _{1'} ^{\star} G^K (1,1')
-G^K (1,1'')\Sigma_\delta(1'', {1'})
%{\delta}(t_{1''} - t_{1'})
= G^K (1,1'') \Sigma^-(1'',1') + G^+ (1,1'') \Sigma^K(1'',1')
\end{equation}
where $s _i =i\partial_{t_i}+\frac{1}{2m}\Delta_i$.
>From the diagonal part we obtain analogous equations  for the 
spectral part of the Green function, $G^A$.

To derive an equal-time formulation 
we perform a Wigner transform of these equations in the space coordinates only. 
For the  Wigner transform
of a general function $F$ we use the notation
\begin{eqnarray}
{f} ({\bf R} , {\bf p}, T ; \tau )  & = &  \int d^3 r e^{-i {\bf p}
{\bf r}}
{F} (
{\bf R} +{{{\bf r}}/{2}} , T + {{\tau}/{2}},
{\bf R} -{{{\bf r}}/{2}} , T - {{\tau}/{2}}
) \, .
\end{eqnarray}
The transforms of  ${{G}}$ and  ${{\Sigma}}$ are denoted
by $g$ and $\sigma$, respectively. 

Introducing the differential operator (Poisson operator)
\begin{eqnarray}
\diamondsuit (f \cdot g) & = & ( {\bf{\bigtriangledown}}_{R} f )
( {\bf{\bigtriangledown}}_{p} g ) -
( {\bf{\bigtriangledown}}_{p} f )
( {\bf{\bigtriangledown}}_{R} g )
\end{eqnarray}
 the equations for the Wigner transformed spectral and kinetic Green functions,
$g^{A}$ and $g^{K}$, are
\begin{eqnarray}
&&\left[ \pm i {{1}\over{2}} {{\partial}\over{\partial T}} + {{1}\over{2m}} 
\left(
{{1}\over{4}} {\bigtriangleup}_{R}  \pm  i {\bf p} {\bf{\bigtriangledown}}_{R}
- {{\bf p}}^2 
\right) + i {{\partial}\over{\partial \tau}}
- e^{ \pm {{i}\over{2}} \diamondsuit }{\sigma}_{\delta}( T \pm {{\tau}/{2}} )
\right]
\left\{
\begin{array}{c}
g^A (T;\tau)\\
g^K (T;\tau)
\end{array}
\right\} =\nonumber \\
&& e^{\pm {{i}\over{2}} \diamondsuit }
\left\{ 
\begin{array}{c}
S^A_{\pm}(T;\tau)\\
S^K_{\pm}(T;\tau) 
\end{array}
\right\}
\end{eqnarray}
with the collision integrals
\begin{eqnarray}
S^A_{\pm} (T;\tau)&=&
\int_{0}^{\tau} dt
{\sigma}^{A}( T \pm  {{t}\over{2}}; \tau - t) \cdot
g^{A} (T \mp {{\tau -t}\over{2}}; t) \\
S^K_{\pm} (T;\tau) &=&
\pm \int_{- \infty}^{\pm \tau} dt
{\sigma}^{A}( T + {{t}\over{2}}; \tau  \mp t) \cdot
g^{K} (T \mp {{\tau \mp t}\over{2}}; \pm t)  \nonumber \\
&&
\mp \int_{-\infty}^{\mp \tau} dt
{\sigma}^{K}( T+ {{t}\over{2}}; \tau \pm t) \cdot
g^{A} (T \pm {{\tau \pm t}\over{2}}; \mp t)
\end{eqnarray}
where we did not indicate the dependence of $g^{A,K}$ and ${\sigma}^{A,K}$
on ${\bf R}$ and ${\bf p}$. The upper and lower signs 
in eqs. (17-19) result when starting from
eqs. (10) and  (11), respectively.
The boundaries of the integrals in eqs. (18,19) come from the step function
in eq. (5).
%In the equations for the spectral component of the Green function, $g^A$, 
%one has to take into account
%the step function in eq. (5) in the boundaries of the integrals.

Setting $\sigma^{>}=\sigma^{<} = 0$, {\it i.e.} when the scattering is
neglected,
the problem is treated in the mean field approximation, the terms on the
right-hand side of the above equations vanish, and eqs. for $g^A$ and $g^K$ are
identical. Therefore one has only  two equations for
$g^{<} = {{1}\over{2}} (g^K -g^A)$ which  is
directly related  to the particle density. In the general case it is neccessary 
to consider all the four equations.
Taking the sum and the difference of  eqs. (17)  we obtain for $g^A$ and $g^K$
two sets of equations
\begin{eqnarray}
&&\left\{
\begin{array}{c}
i{{\partial}\over{\partial T}} + i {{\bf p}\over{m}}  
{\bf{\bigtriangledown}}_{R}-
\left[
e^{{{i}\over{2}} \diamondsuit }{\sigma}_{\delta}( T + {{\tau}/{2}} ) -
e^{ - {{i}\over{2}} \diamondsuit }
{\sigma}_{\delta}( T - {{\tau}/{2}} )
\right] \\
{{1}\over{m}} (
{{1}\over{4}} {\bigtriangleup}_{R}  - {\bf p}^2 ) + 
2 i {{\partial}\over{\partial \tau}} -
\left[
e^{{{i}\over{2}} \diamondsuit }{\sigma}_{\delta}( T + {{\tau}/{2}} ) +
e^{ - {{i}\over{2}} \diamondsuit }
{\sigma}_{\delta}( T - {{\tau}/{2}} )
\right]
\end{array}
\right\}
g^{A,K}(T;\tau) = \nonumber \\
&&\left\{
\begin{array}{c}
e^{{{i}\over{2}} \diamondsuit }S_+^{A,K}(T;\tau) - 
e^{-{{i}\over{2}} \diamondsuit }S_-^{A,K}(T;\tau)\\
e^{{{i}\over{2}} \diamondsuit }S_+^{A,K}(T;\tau) + 
e^{-{{i}\over{2}} \diamondsuit }S_-^{A,K}(T;\tau)
\end{array}
\right\}
\end{eqnarray}
The  equations in the upper row are called  kinetic equations because they
involve the time  evolution in the global time T. The ones in the lower row 
are called constraint equations \cite{ETH}, since
under certain conditions (see below) they are algebraic equations.

%Instead of taking $-\infty$ as the lower boundary of the Schwinger-Keldysh
%contour it might be advantageous to choose the lower boundary 
%to be $T_0$, {\it e.g.} if
%initial conditions are given at a certain time $T_0$. This implies that the 
%lower 
%limit in the integrals in the equations (19)  must be
%changed, namely
%$-\infty$ in $S^K_{\pm}$   must be replaced by 
%$ \frac{\tau}{2}-(T-T_0)$
%resp. $-\frac{\tau}{2}-(T-T_0)$.
%However,
%if $(T-T_0)$ is larger than a typical correlation
%time of the system, the lower limit of the integrals can be
%restored to $-\infty$.
%The  limits of the integral in the equations for the spectral part 
%are independent of the initial time $T_0$.

\vskip 1truecm
\section{Hierarchies of equations for the spectral functions}

The eqs. (20) are, of course, still exact. The quantities  
$g^{A,K}(T;\tau)$,
 $\sigma^{A,K}(T;\tau)$ depend on two time arguments, the global time 
$T$ and the relative time $\tau$. The integrations in the 
collision terms, (eqs. (18) and (19)),
involve these functions for all relative times from  $-\infty$ to 
$+\infty$.
Thus these equations are extremely difficult to handle in practice. 
A drastic approximation is to put the relative time equal to zero, 
{\it i.e.} to write an equal-time formalism. This route has been 
followed in other approaches \cite{QP}. This
is equivalent to the so-called quasi-particle approximation,  in which 
the energy dependence of the spectral function is approximated by a 
$\delta$ function,
%a relation $\omega ({\bf p})$ is fixed,  
and thus off-shell properties of the
distribution functions are neglected. As mentioned in  the introduction this
is a drastic approximation for many reasons.  An improvement
is therefore expected by expanding all quantities around $\tau = 0$. The
off-shell information is then contained in the time derivatives with respect
to $\tau$ at $\tau=0$, either 
inside of the convergence radius by Taylor expansion or by analytic
continuation. We therefore will develop  an infinite set of equations, or a
hierarchy of equations, for these equal-time derivatives originating from the
kinetic and the constraint equations.

The  equal-time $\tau$-derivatives of order $n$ $g ^{(n)} (T, \tau)$
are  related
to the energy moments of order $n$ of the Fourier transformed function
$g(T,\omega)$
since we can write
\begin{eqnarray}
g^{(n)}(T)&:=&\partial_\tau^n g(T,\tau) |_{\tau=0}\nonumber\\
&=&\partial_\tau^n \int \frac{d\omega}{2\pi} e^{i\omega \tau}g(T,\omega)
\Big|_{\tau=0}\nonumber\\
&=&\frac{(-i)^n}{2\pi}\int d\omega\,\omega^n g(T,\omega) \,.
\end{eqnarray}
However,  the energy moments may
not exist in general. {\it E.g.} if $g(T,\omega)$ behaves like 
$\frac{1}{\omega^2}$ for large
$\omega$ , then the second and higher energy
moments of the spectral component diverge. Such a behaviour {\it e.g.} occurs 
in the gradient approximation of the spectral function 
neglecting the dependence of the
imaginary part of the retarded self-energy on $\omega$.
The reason  is that  integration and differentiation cannot,
in general,  be
interchanged in  eq. (21). In contrast to the energy moments
the time derivatives $g^{(n)}(T)$ exist as seen from the definition of $g$ in 
terms of operator products.

The problem of the divergence of the energy moments could be overcome  by 
introducing a cut-off in  the energy-integral, but we do not choose this
procedure here. However, we can  retain the physical intuition that,
{\it e.g.} for the spectral Green function,
 the zeroth
 derivative $g^{(0)}(T)$ is related to the normalization of the function, the
first derivative $g^{(1)}(T)$ to the mean energy of the quasi-particle, and
the second derivative $g^{(2)}(T)$ to the width of the spectral function.
Indeed, neglecting the possible divergence problems, we have for the spectral
function $A$ defined {\it e.g.} by  Henning
\cite{THI} $A(T, \omega) = - {{1}\over{2 \pi i}}\int d \tau e^{i \omega \tau}
g ^A(T; \tau) $. Then, $ \int d\omega \omega ^n A(T,\omega) = i^{n+1}
g^{A(n)}  $ and the $n ^{th}$ $\tau$-derivative is  exactly the $n ^{th}$
energy moment of the four dimensional Wigner transform of the spectral
function up to a factor  $i^{n+1}$.

We start by  setting up the hierarchy for the spectral
function $g^A$. From the  upper row of eqs. (20) for $g^A$
we obtain the hierarchy of kinetic equations for
the  equal-time $\tau$-derivatives of the spectral component. The equal-time
commutation relation, through eq.(9), results into
\begin{eqnarray}
g^{A(0)} & = & - i
\end{eqnarray}
which is  the sum rule for the spectral  function.
Using  this relation, the first three equations of the kinetic hierarchy are
\begin{eqnarray}
&&i  {{\partial}\over{\partial T}} g^{A(0)} -2 i \sin {{1}\over{2}}
\diamondsuit ( \epsilon \cdot  g^{A(0)} )  =  0
\nonumber \\
&&i  {{\partial}\over{\partial T}} g^{A(1)} -2 i \sin {{1}\over{2}}
\diamondsuit (\epsilon  \cdot  g^{A(1)} ) -\cos  {{1}\over{2}}
\diamondsuit (  {{\partial \epsilon }\over{\partial T}} \cdot  g^{A(0)} )
  =  0
\nonumber \\
&&i  {{\partial}\over{\partial T}} g^{A(2)} -2 i \sin {{1}\over{2}}
\diamondsuit ( \epsilon \cdot  g^{A(2)} ) -2 \cos  {{1}\over{2}}
\diamondsuit (  {{\partial \epsilon }\over{\partial T}} \cdot  g^{A(1)} ) =
\nonumber \\
&&
 2 i \sin {{1}\over{2}}
\diamondsuit ( \sigma^{A(0)} \cdot  g^{A(1)} ) -
 i  {{\partial \sigma^{A(0)} }\over{\partial T}}
\end{eqnarray}
where $\epsilon ={{{\bf p}^2}\over{2m}} + \sigma_{\delta}(T)$ and where
we used
the  identities
\begin{eqnarray}
{{\bf p}\over{m}}  {\bf{\bigtriangledown}}_{R}f & = & -2 \sin {{1}\over{2}}
\diamondsuit ( {{{\bf p}^2}\over{2m}} \cdot  f )  \nonumber \\
{{1}\over{m}} (
{{1}\over{4}} {\bigtriangleup}_{R}  - {\bf p}^2 )f & = &
-2 \cos {{1}\over{2}}\diamondsuit ( {{{\bf p}^2}\over{2m}} \cdot  f )  \nonumber
\end{eqnarray}
The integrals on the rhs of eqs. (20) for $g^A$
vanish at $\tau =0$, eqs. (18).
>From the second row of equation (20) we obtain a constraint hierarchy  
from which we write down 
the first
two members.
\begin{eqnarray}
&& -2  \cos {{1}\over{2}}
\diamondsuit ( \epsilon \cdot  g^{A(0)} ) + 2 i g^{A(1)}  =  0
\nonumber \\
&& -2  \cos {{1}\over{2}}
\diamondsuit ( \epsilon \cdot  g^{A(1)} ) + 2 i g^{A(2)}  =
2 \cos  {{1}\over{2}}
\diamondsuit (  \sigma^{A(0)}  \cdot  g^{A(0)} )
\end{eqnarray}

In general the $n^{th}$ equation of the constraint hierarchy gives the $n^{th}$
$\tau$-derivative of the spectral function, $g^{A(n)}$,
in terms of the lower order $\tau$-derivatives of the spectral function, 
$g^{A(j)}$ 
and  of the spectral
self-energy , $\sigma ^{A(j)}$ with $j \le n-1$. On the
other hand the  $(n+1)^{th}$  equation of the kinetic hierarchy is a kinetic
equation for the $n^{th}$ $\tau$-derivative of the spectral function, $g^{A(n)}$,
in which only the lower order $\tau$-derivatives of the spectral function and
spectral self-energy, $g^{A(j)}$ and  $\sigma ^{A(j)}$ for $j \le n-1$,   
appear.
The first kinetic equation (23) is satisfied by $g^{A(0)} = - i$, eq. (22).

The natural question arises: are the $n^{th}$ equation of the constraint
hierarchy and the $(n+1)^{th}$ equation of the kinetic hierarchy  compatible?
In fact, the $(n+1)^{th}$ kinetic equation is not a new equation, but it
can be expressed as a  combination of  lower order constraint and kinetic
equations. 
Since we have not made any approximation yet,
this result is  true beyond the mean field approximation,  in
which only the singular part of the self-energy, {\it i.e.} Hartree-Fock 
self-energy, is retained. We will show
this  explicitely in the following up to order $n=2$. This means 
for the spectral part
that the algebraic
solutions of the constraint hierarchy satisfy the kinetic hierarchy,
and that the
time evolution is  described only by the $T$-dependence   
of the self-energies.

With eq. (22) and eq. (24) we have from the constraint hierarchy
\begin{eqnarray}
 g^{A(1)} & = & - \epsilon \\
 g^{A(2)} & = &  i \cos  {{1}\over{2}}\diamondsuit
(  \epsilon  \cdot  \epsilon ) - \sigma^{A(0)}
\end{eqnarray}
%It may seems that the constraint hierarchy is  decoupled
%from the hierarchies for the  kinetic part of the Green function, $g^{K}$, but
%we note that the self-energy depends on $g^{K}$.

To prove the compatibility of the kinetic and constraint hierarchies we have to
verify that the solutions (25,26) satisfy eqs. (23). This is done in Appendix B
using the properties of the differential operator $\diamondsuit$ given 
in Appendix A. 
Therefore, only the constraint hierarchy is necessary, the kinetic hierarchy
being redundant, as the solutions of the former  satisfy automatically
the latter one. The $\tau$-derivatives of the spectral function 
can be determined
from the constraint hierarchy {\it if} the  $\tau$-derivatives
of the spectral
self-energy are known. In general, these depend on the  $\tau$-derivatives of the
kinetic part of the Green function,  $g^{K}$, as will be shown later .
Thus the spectral and kinetic Green functions are coupled.

\vskip 1truecm
\section{Hierarchies of equations for the kinetic  Green functions}

Following the same way as for the spectral function we obtain
from eqs. (20) two infinite
hierarchies of equations for the equal-time $\tau$-derivatives of the 
kinetic part
$g^{K}$. 
Now also the equal-time derivatives of the collision integral enter.
The first three equations of the kinetic hierarchy are
\begin{eqnarray}
&&i  {{\partial}\over{\partial T}} g^{K(0)} -2 i \sin {{1}\over{2}}
\diamondsuit ( \epsilon \cdot  g^{K(0)} )  =
e^{{{i}\over{2}} \diamondsuit } S_{+}^{K(0)}  -
e^{ - {{i}\over{2}} \diamondsuit } S_{-}^{K(0)}  \nonumber \\
&&i  {{\partial}\over{\partial T}} g^{K(1)} -2 i \sin {{1}\over{2}}
\diamondsuit (\epsilon  \cdot  g^{K(1)} ) -
\cos  {{1}\over{2}}
\diamondsuit (  {{\partial \epsilon }\over{\partial T}} \cdot  g^{K(0)} ) =     
\nonumber \\ &&
e^{{{i}\over{2}} \diamondsuit }  S_{+}^{K(1)} -
e^{ - {{i}\over{2}} \diamondsuit }  S_{-}^{K(1)}
 \nonumber \\
&&i  {{\partial}\over{\partial T}} g^{K(2)} -2 i \sin {{1}\over{2}}
\diamondsuit ( \epsilon \cdot  g^{K(2)} )
-2 \cos  {{1}\over{2}}
\diamondsuit (  {{\partial \epsilon }\over{\partial T}} \cdot  g^{K(1)} ) -
{{i}\over{2}} \sin  {{1}\over{2}}
\diamondsuit (  {{{\partial}^2 \epsilon }\over{\partial T ^2}} \cdot  g^{K(0)} )
 =  \nonumber \\ &&
e^{{{i}\over{2}} \diamondsuit }  S_{+}^{K(2)} -
e^{ - {{i}\over{2}} \diamondsuit }  S_{-}^{K(2)}  
\end{eqnarray}
where $S_{\pm}^K$ are given by eq. (19).

The first two equations of the constraint hierarchy are
\begin{eqnarray}
&& -2  \cos {{1}\over{2}}
\diamondsuit ( \epsilon \cdot  g^{K(0)} ) + 2 i g^{K(1)}  =  e^{{{i}\over{2}} 
\diamondsuit } S_{+}^{K(0)} +
e^{ - {{i}\over{2}} \diamondsuit } S_{-}^{K(0)} \nonumber
\end{eqnarray}
\begin{equation}
-i \sin  {{1}\over{2}}
\diamondsuit (  {{\partial \epsilon }\over{\partial T}} \cdot  g^{K(0)} )
-2  \cos {{1}\over{2}}
\diamondsuit ( \epsilon \cdot  g^{K(1)} ) + 2 i g^{K(2)}  =
e^{{{i}\over{2}} \diamondsuit }  S_{+}^{K(1)}  +
e^{ - {{i}\over{2}} \diamondsuit }  S_{-}^{K(1)}
\end{equation}

Let us first consider  the mean
field approximation, when
  collisions are neglected ($ S_{\pm}^K(\tau ) \equiv 0$). 
The $n^{th}$ equation of the constraint hierarchy gives
the $n^{th}$  $\tau$-derivative,  $g^{K(n)}$, in terms of the lower order
 $\tau$-derivatives,
$g^{K(j)}$ for $j \le n-1$. The $(n+1)^{th}$ equation of the kinetic hierarchy
is an evolution equation for the $n^{th}$  $\tau$-derivative, $g^{K(n)}$.
In this case
we  show in Appendix B that at least up to order $n=2$ 
only the first equation of the
kinetic hierarchy and all the equations of the constraint hierarchy are
independent; the $n^{th}$ equation of the kinetic hierarchy, $n>1$,
can be expressed in terms of the lower order  kinetic and constraint equations.
Therefore the two hierarchies are compatible in the mean field approximation 
(at least up to order $n=2$).
 We note that   in general
the singular part of the self-energy on the contour
may depend on ${\bf p}$, in contrast to the usual assumption
\cite{ETH}.

The inclusion of the collision contribution is more difficult because
we have to approximate the integrals 
$S_{\pm}^{K(0)} $
and $S_{\pm}^{K(1)}$. The main
contribution to the integrals
\begin{eqnarray}
S_{\pm}^{K(0)}   =   \pm
\int_{-\infty}^0\, dt [\sigma^A(T+\frac{t}{2}, \mp t) g^K(T+\frac{t}{2}, \pm t)
- \sigma^K(T+\frac{t}{2}, \pm t) g^A(T+\frac{t}{2},\mp t)] \nonumber
\end{eqnarray}
 comes from small values of the variable of integration $t$ since  large 
 values  
involve  Green functions at large relative time, which are expected
to be small. Therefore we can approximate the integrals using as 
a lower limit a 
judiciously choosen parameter
$-\Delta$ instead of $-\infty$ (or $T_0-T$).
For this purpose we expand
the functions under the integral of the type $f(T+t/2,\pm t)$ in
$t$
around $f(T,t=0)$, i.e. around the equal-time value at current göobal time
$T$,  and integrate over $dt$.
If we expand the integrand to first order in $t$
we obtain the second order approximation in the parameter $\Delta$
for the collision integrals
\begin{eqnarray}
S_{\pm}^{K(0)}  &=&\pm  \Delta
\left(
1- {{\Delta}\over{4}}{{\partial}\over{\partial T}}
\right)
[ \sigma ^{A(0)} g ^{K(0)} - \sigma ^{K(0)} g ^{A(0)} ]
 \nonumber\\ &&
+ {{1}\over{2}} \Delta ^2 (\sigma^{A(1)} g^{K(0)}-\sigma^{A(0)} g^{K(1)}
-\sigma^{K(1)} g^{A(0)}+\sigma^{K(0)} g^{A(1)})
\Big]
\,,
\end{eqnarray}
where the time argument $T$ has been omitted everywhere. 
For the derivatives of 
$S_{+}^K$ and $S_{-}^K$, 
by expanding the integrand in zero order in $t$,
we obtain an expression linear in $\Delta$
\begin{eqnarray}
 S_{\pm}^{K(1)} &=&
\frac{1}{2}
(\sigma^{A(0)} g^{K(0)}+\sigma^{K(0)} g^{A(0)} )\nonumber\\
&
\pm 
&
%\bigg\{   
{{ \Delta  }\over{2}}
\bigg[
\Big(
\sigma ^{A(1)} g ^{K(0)} + \sigma ^{A(0)} g ^{K(1)}
-\sigma ^{K(1)} g ^{A(0)} - \sigma ^{K(0)} g ^{A(1)}
\Big)
 \nonumber\\
&
\pm 
&
\frac{1}{2}
\bigg(
{{\partial \sigma ^{A(0)} }\over{\partial T}}   g ^{K(0)}
- \sigma ^{A(0)}  {{\partial g ^{K(0)} }\over{\partial T}}
- {{\partial \sigma ^{K(0)} }\over{\partial T}}   g ^{A(0)}
+ \sigma ^{K(0)}  {{\partial g ^{A(0)} }\over{\partial T}}
\bigg)
\bigg] 
%\nonumber \\ &
%-  \frac{ \Delta  ^2}{4}   \bigg[
%\pm \frac{1}{4}
%\bigg(
%{{\partial ^2 \sigma ^{A(0)} }\over{\partial T ^2}} g ^{K(0)}
%- \sigma ^{A(0)} {{\partial ^2 g ^{K(0)} }\over{\partial T ^2}}
%-{{\partial ^2 \sigma ^{K(0)} }\over{\partial T ^2}} g ^{A(0)}
%+ \sigma ^{K(0)} {{\partial ^2 g ^{A(0)} }\over{\partial T ^2}}
%\bigg)  \nonumber\\
%&
%+\bigg(
%{{\partial \sigma ^{A(0)} }\over{\partial T }} g ^{K(1)}
%+ \sigma ^{A(1)} {{\partial  g ^{K(0)} }\over{\partial T }}
%-{{\partial  \sigma ^{K(0)} }\over{\partial T }} g ^{A(1)}
%- \sigma ^{K(1)} {{\partial  g ^{A(0)} }\over{\partial T }}
%\bigg) \nonumber\\
%&
%&\mp \Big(
% \sigma ^{A(2)}  g ^{K(0)} -\sigma ^{A(0)}  g ^{K(2)}
%-\sigma ^{K(2)}  g ^{A(0)} +\sigma ^{K(0)}  g ^{A(2)}
%\Big)
%\bigg] 
\end{eqnarray}
Another  way to approximate the
collision integrals by a convergence factor $e^{-\alpha}$ is presented 
in the appendix C. There it is seen  that the results are  
essentially  the same
as above, with $\Delta $  replaced by  
a term proportional to $1/\alpha$.

Using the above expressions (29) and (30)  
for the collision intregrals
the first two equations of  the
kinetic and constraint hierarchies
for the kinetic Green function,
$g ^K$, include memory effects due to the presence of time
derivatives.  The eqs. (29) and  (30) are second, respectively first, order
polynomial approximations  in $\Delta $ for the collision terms. In
general, in the
$k^{th}$ order approximation of $S_{\pm}^{K(0)}$ in $\Delta$
 appear   $\tau$-derivatives of the Green functions
and of the self-energy
up to order $k-1$.
As we will see in the next section, the latter can be obtained  in terms
of only the same order  $\tau$-derivatives of the Green functions. However, 
in the same
approximation in $\Delta $, the collision terms in the $n^{th}$-order 
equations of the two
hierarchies contain  $\tau$-derivatives of the self-energy up to order $n+k-2$.
Thus the appropriate truncation of the hierarchies
is not as obvious as in the mean field approximation; only the
hierarchy for the spectral function truncates by itself, completely
similar to the mean field case.

If we approximate the collision terms $S_{\pm}^{K(0)}$ in order $k$, $k \ge 1$,
with respect to $\Delta$
then the first  equation of the kinetic hierarchy contains  the 
$\tau$-derivatives of the Green functions up to order $k-1$. 
Therefore in general 
we choose to 
approximate the collision terms  which appear in 
the $n^{th}$-order equations of the two hierarchies, $S_{\pm}^{K(n-1)}$,
in  order $k-(n-1)$.
Within this approximation the 
first $k-1$ equations of the constraint hierarchy contain also the  
$\tau$-derivatives of  the Green functions up to order $k-1$. Therefore, one
has $k$ equations for $k$  $\tau$-derivatives. The  $\tau$-derivatives of
the self-energies  appear up
to order $k-1$ , {\it i.e.} $N$  $\tau$-derivatives and they  can be
obtained self-consistently.

In the following we will consider  the $k=2$ approximation. The kinetic and 
constraint equations are
\begin{eqnarray}
&&i  {{\partial}\over{\partial T}} g^{K(0)} -2 i \sin {{1}\over{2}}
\diamondsuit ( \epsilon \cdot  g^{K(0)} )  =   \nonumber \\&&
 2 \Delta \left(1 - {{\Delta}\over{4}}   {{\partial}\over{\partial T}}\right) 
\cos {{1}\over{2}}
\diamondsuit
( \sigma ^{A(0)} g ^{K(0)} - \sigma ^{K(0)} g ^{A(0)} )
 \nonumber \\ &&
+  i     \Delta  ^2
 \sin {{1}\over{2}}
\diamondsuit\left(  \sigma ^{A(1)} g ^{K(0)} - \sigma ^{A(0)} g ^{K(1)}
- \sigma ^{K(1)} g ^{A(0)}  + \sigma ^{K(0)} g ^{A(1)}
\right)
 \\&&
 -2  \cos {{1}\over{2}}
\diamondsuit ( \epsilon \cdot  g^{K(0)} ) + 2 i g^{K(1)}  =  \nonumber \\&&
    2i \Delta 
 \left(1 - {{\Delta  }\over{4}}   
   {{\partial}\over{\partial T}}\right) 
\sin {{1}\over{2}}
\diamondsuit
( \sigma ^{A(0)} g ^{K(0)} - \sigma ^{K(0)} g ^{A(0)} )
 \nonumber \\ &&
+ \Delta  ^2
\cos {{1}\over{2}}\diamondsuit  
\left(  \sigma ^{A(1)} g ^{K(0)} - \sigma ^{A(0)} g ^{K(1)}
- \sigma ^{K(1)} g ^{A(0)}  + \sigma ^{K(0)} g ^{A(1)}
\right)
\end{eqnarray}

The familiar Vlasov and Boltzmann equations are particular cases of the
above approximation.
When in the collision integrals $S_{\pm}^K$  only the terms that are
zeroth order in $\Delta $ are retained (the collision integrals are neglected),
the equations for $g^{K(0)}$ has the same form as the corresponding equation
for  $g^{A(0)}$. Thus the first equation of the kinetic hierarchy is
\begin{eqnarray}
i  {{\partial}\over{\partial T}} g^{K(0)} -2 i \sin {{1}\over{2}}
\diamondsuit ( \epsilon \cdot  g^{K(0)} ) & = &  0 \nonumber
\end{eqnarray}
which in the gradient approximation is the Vlasov equation with a self 
consistent mean-field,
which is included in $\epsilon$.
The validity of this approximation is
not directly related to  the quasi-particle approximation. Here we only use 
the assumption that
the collision term can be approximated in zeroth order
in $\Delta $.

The lowest order nontrivial approximation is $k=1$, which results in
 taking only 
terms with   $g^{A(0)}$ and  $g^{K(0)}$ in eqs. (31,32). We then have the
kinetic equation for $g^{K(0)}$
\begin{eqnarray}
i  {{\partial}\over{\partial T}} g^{K(0)} -2 i \sin {{1}\over{2}}
\diamondsuit ( \epsilon \cdot  g^{K(0)} ) & = &
2 \Delta  \cos {{1}\over{2}} \diamondsuit (\sigma ^{A(0)} g ^{K(0)} -
\sigma ^{K(0)} g ^{A(0)} ) \, .
\end{eqnarray}
Let us note that the term $(\sigma ^{A(0)} g ^{K(0)} -
\sigma ^{K(0)} g ^{A(0)} )$ in the above equation can be put into the
 familiar form of the collision term of a Boltzmann-like equation, {\it i.e.}
$$
(\sigma ^{A(0)} g ^{K(0)} -
\sigma ^{K(0)} g ^{A(0)} )=
(\sigma ^{>(0)} - \sigma ^{<(0)})(g ^{>(0)} + g ^{<(0)}) -
(\sigma ^{>(0)} + \sigma ^{<(0)})(g ^{>(0)} - g ^{<(0)})
$$
\begin{equation}
=2(\sigma ^{>(0)} g ^{<(0)} - \sigma ^{<(0)} g ^{>(0)})
\end{equation}
 Eq. (33) goes beyond to the gradient approximation, and
the higher order space derivatives contain information about
quantum effects. In the gradient approximation we obtain a Boltzmann-like
equation for the zeroth $\tau$-derivative of the Green function.

Consider now the k=2 approximation for the collision integrals, eqs.(31, 32),
we need to calculate  the first two  $\tau$-derivatives of the kinetic Green
function.  
%Only in the gradient
%approximation  the constraint equation (32) gives the first   
%$\tau$-derivative
%in terms of zeroth one, otherwise it is a differential equation.
In the gradient approximation  ($\cos {{1}\over{2}}\diamondsuit  \to 1$,
$\sin {{1}\over{2}}\diamondsuit  \to  {{1}\over{2}}\{, \}$)  eq. (32) is
 an expression for    $g^{K(1)}$ depending on $g^{K(0)}$ and the 
first   order
$\tau$-derivative
of the self-energy. This expression can be substituted into  the kinetic 
equation (31) yielding
 a transport equation for  $g^{K(0)}$.
Since $g^{K(1)}$ appears in  eq. (31) under the
Poisson bracket, the Poisson bracket in  eq. (32) has to be neglected
consistently with the gradient approximation, because a substitution of it
would  result in a
term in  eq. (31) which involves second order derivatives. 
Moreover, $g^{K(1)}$ appears in eq. (31) in the second order term in $\Delta$.
Therefore we can use in eq. (31)  the solution of
the algebraic eq. (32) for $g^{K(1)}$ in zeroth order in $\Delta$  
 $g^{K(1)}  \simeq  -i
\epsilon g^{K(0)} $.
%\left\{ 
%(1+  i{{  \Delta  ^2}\over{2}}  \sigma ^{A(0)})
%\epsilon g^{K(0)} 
%+  {{ \Delta  ^2}\over{2}}
%\left(
% \sigma ^{A(1)} g ^{K(0)} +i \sigma ^{K(1)} -\epsilon  \sigma ^{K(0)}
%\right)
%\right\} 
%\,, \nonumber
Thus, in a
consistent  approximation,  substituting this expression in  eq. (31),
we obtain $$
i  {{\partial}\over{\partial T}} g^{K(0)} - i\{ \epsilon , g^{K(0)} \}
-i {{\Delta ^2}\over{2}} 
\left[
\{ \sigma ^{A(1)} , g^{K(0)} \} + \epsilon \{ i \sigma ^{A(0)} , g^{K(0)} \}
\right]
 =   $$
\begin{equation}
{2 \Delta } 
\left(1 - {{\Delta  }\over{4}}   {{\partial}\over{\partial T}}\right)
( \sigma ^{A(0)} g ^{K(0)} - \sigma ^{K(0)} g ^{A(0)} )
+ i {{\Delta ^2}\over{2}} 
\left[
\{ i \sigma ^{A(0)} , \epsilon \} g^{K(0)} 
- \{ i \sigma ^{K(0)}  , \epsilon\} g^{A(0)} 
\right]
\end{equation}
where $\sigma ^{A(1)}$ can be calculated using the same 
approximation in  for $g^{K(1)}$ in terms of $g^{K(0)}$ as above.

%This equation can be written in a Boltzmann-like form
%\begin{displaymath}
%\left( 1- i \Delta  ^2  \sigma ^{A(0)} \right)   {{\partial}\over{\partial
%T}}  g^{K(0)} -
%\left\{  \epsilon -{{\Delta  ^2}\over{2}}
%%\left(
%%\sigma ^{A(1)} 
%i \epsilon \sigma ^{A(0)}
%%\right)
%,g^{K(0)}\right\}  =
%\end{displaymath}
%\begin{equation}
% -2 \Delta 
%  \left[ g^{K(0)}
%\left[
%\left( 1 - {{\Delta }\over{4}}{{\partial}\over{\partial T}} \right)
%{{i \sigma ^{A(0)}}} - {{\Delta }\over{4}}
%\left\{ {{i \sigma ^{A(0)}}} ,
%\epsilon \right\}
%\right]
%- g^{A(0)}
%\left[
%\left( 1 - {{\Delta }\over{4}}{{\partial}\over{\partial T}}  \right)
%{{i \sigma ^{K(0)}}}
%\right]
%\right] 
%\end{equation}

%The first term has a renormalization factor in front of the time derivative
%and the next Poisson bracket  involves  the expression
%$$
%\epsilon -{{\Delta  ^2}\over{2}}
%%\left(
%%\sigma ^{A(1)} 
%i \epsilon \sigma ^{A(0)}
%%\right)
%$$
%which reduces to the Hartree-Fock
%Hamiltonian $\epsilon$ when $\sigma ^{A(0)}$ is neglected. 
%The collision term can be written in the usual form, eq. (34), 
%if we make
%the identifications
%\begin{eqnarray}
%\tilde{\sigma} ^> & = &
%\left( 1 - {{\Delta }\over{4}}{{\partial}\over{\partial T}} \right)
%{{ \sigma ^{>(0)}}} - {{\Delta }\over{8}}
%\left\{ {{ \sigma ^{A(0)}}} ,
%\epsilon \right\} \nonumber \\
%\tilde{\sigma} ^< & = &
%\left( 1 - {{\Delta }\over{4}}{{\partial}\over{\partial T}} \right)
%{{ \sigma ^{<(0)}}} + {{\Delta }\over{8}}
%\left\{ {{ \sigma ^{A(0)}}} ,
%\epsilon  \right\} \nonumber
%\end{eqnarray}

To go beyond the gradient approximation one has to 
keep the higher order gradient terms in the otherwise algebraic constraint
equation 
for $g ^{K(1)}$.

%\newpage

\vskip 1truecm
\section{Self-energy in the Born approximation}

To solve the equations for the  $\tau$-derivatives of the Green functions  
we need the corresponding  $\tau$-derivatives of the
self-energies. The self-consistent Hartee-Fock self-energy is singular
in time on the
contour; it appears in the equations as a potential $\sigma _{\delta}$ 
(momentum dependent in general) and it depends only on the zeroth 
$\tau$-derivative of the Green functions.
The Hartree self-energy can be written \cite{DEC}
\begin{eqnarray}
\Sigma_{H}(1,1') & = & -i\delta (t_1 - t_{1'} )
 \delta ({\bf x}_1 - {\bf x}_{1'} )
\int d^3 x_2 V ({\bf x}_1 - {\bf x}_{2} )
G^< ({\bf x}_2 , t_1  , {\bf x}_{2} , t_1  ) \nonumber
\end{eqnarray}
The  Wigner transformation  does not depend on momentum  due
to the presence of the $\delta$-function in space coordinates and it also 
has no $\tau$-dependence.
One obtains
\begin{eqnarray}
\sigma_{H}({\bf R}, T) & = & -i \int d^3 \xi V ({\bf \xi})
{{1}\over{(2\pi )^3}} \int d^3 p  g^{<(0)} ({\bf R} -{\bf \xi}, {\bf p} , T)
\nonumber \\
& = & - {{i}\over{(2\pi )^3}} \sum _m {{1}\over{m!}} {\partial}_R ^m
\left[\int d ^3 p g^{<(0)} ({\bf R} , {\bf p} , T) \right]
\int  d^3 \xi  {\bf \xi}^m V ({\bf \xi})
\end{eqnarray}
The Green function appearing above is
\begin{eqnarray}
 g^{<(0)} ({\bf R} , {\bf p} , T)& = & - {{i}\over{2}}
\left(
i  g^{K(0)} ({\bf R} , {\bf p} , T) -1
\right)
\end{eqnarray}
 using definitions (7) and the
sum rule (22) for the spectral function.
The gradient expansion in the second line of eq. (36) gives
 non-locality contributions to the Hartree self-energy due to
the dependence of the Green function  $g^{K(0)}$ on ${\bf R}$
 which are not present in a translationally  invariant system.
However, the lowest non-locality contribution
 (proportional to the gradient of the Green function)
has a factor
$\int  d^3 \xi  {\bf \xi} V ({\bf \xi})$ which is zero for
inter-particle interaction potentials with
$V ({\bf \xi}) = V ( - {\bf \xi})$. Therefore, the specification  of
the Hartree self-energy in the local approximation
\begin{eqnarray}
\sigma_{H}({\bf R}, T) & = &- {{i}\over{(2\pi )^3}} \left[\int d ^3 p g^{<(0)} 
({\bf R} , {\bf p} , T) \right] \left[
\int  d^3 \xi   V ({\bf \xi}) \right]
\end{eqnarray}
is justified at least in the gradient approximation. The
second order term in the gradient expansion in eq. (36)   for spherically
symmetric potentials  $V ({\xi})$ is proportional to
\begin{eqnarray}
\bigtriangleup _R \left[\int d ^3 p g^{<(0)} ({\bf R} , {\bf p} , T) \right]
\int  d ^3 \xi  {\xi}^4  V ({\xi}) \, .
\end{eqnarray}
 For a $\delta$-interaction
potential the local approximation is exact; as the nucleon - nucleon potential
is of short range, one can expect that in nuclear processes the local 
approximation is reasonable
even for large non-localities. 
%However, when the
%interaction range in large, the above term cannot be neglected even for
%a smooth ${\bf R}$-dependence . 
%As we are interested mainly in
%the description of nuclear collisions, we restrict ourselves in the following
%to the local approximation.

In the same manner we obtain for  the Fock  self-energy  \cite{DEC}
\begin{eqnarray}
\sigma_{F}({\bf R}, {\bf p} , T) & = & {{i}\over{(2\pi )^3}}
\int d^3 \xi  e^{-i {\bf p} {\bf \xi} } V ({\bf \xi})
 \int d^3 p'  g^{<(0)} ({\bf R} , {\bf p}' , T) e^{-i  {\bf p}' {\bf \xi} }
\nonumber \\
& = & {{i}\over{(2\pi )^3}} \int d^3 p'  g^{<(0)} ({\bf R} , {\bf p'} , T)
v ({\bf p}'-{\bf p}) \, ,
\end{eqnarray}
where $v$ is the Fourier transform of the interaction potential.
%For a $\delta$-interaction
%potential the sum of the Hartree and Fock self-energies is zero (there is no
%mean field if the interaction is local and  fourth order polynomial in
%the fermionic fields).
The above expresion  has no gradient corrections.
Thus, the Fock contribution is the convolution of the Green function $g^{<(0)}$ 
with the Fourier transform
of the interaction potential.
We note that the Hartree-Fock self-energy can be obtained from only zero order  
$\tau$-derivatives of the
Green function $g^{K(0)}$ using eq. (37).

The lowest order contribution to the non-singular part of the self-energy on 
the Schwinger-Keldysh contour is given by the  direct and exchange
Born diagrams \cite{DEC}. We will discuss here only the contribution for the 
direct
term, the exchange term is given in an analogous way.
\begin{eqnarray}
&&\sigma_{Bd}^{>,<}({\bf R}, {\bf p} , T ; \tau)  =  {{1}\over{(2\pi )^9}}
\int d^3 \xi  e^{-i  {\bf p} {\bf \xi} }
\int d^3  {\rho}_1 \int d^3  {\rho}_2  V ({\bf \rho}_1) V ({\bf \rho}_2)
 \int d^3 p_1  e^{+i  {\bf p}_1 {\bf \xi} } \nonumber \\ & &
g^{>,<} ({\bf R} , {\bf p}_1 , T; \tau)
 \int d^3 p_2  e^{i  {\bf p}_2
({\bf \xi} -({\bf \rho}_1 + {\bf \rho}_2)) }
g^{>,<} ({\bf R} + {{{\bf \rho}_2 - {\bf \rho}_1}\over{2}}, {\bf p}_2, T;\tau)
\nonumber \\ & &
 \int d^3 p_3  e^{-i  {\bf p}_3
({\bf \xi}-({\bf \rho}_1 + {\bf \rho}_2)) }
g^{<,>} ({\bf R} + {{{\bf \rho}_2 - {\bf \rho}_1}\over{2}}, {\bf p}_3, T; -\tau)
\end{eqnarray}
Expanding eq. (41) around $\tau =0$, it is seen  that the $n^{th}$  
$\tau$-derivative of the Born
self-energy  involves the  $\tau$-derivatives of the
Green functions up to order $n$.
%%%%%%%%%%%%%%%%%%%%%%%%%
This property remains true if one replaces the Born approximation for the 
self-energy
by the T-matrix approximation if the off-shell values of the T-matrix can be 
approximated by the on-shell ones as it was discussed in \cite{cro}.
%%%%%%%%%%%%%%%%%%%%%%%%%%%%%
Further we make a gradient expansion in eq. (41)
of  the last two Green functions 
around ${\bf R}$.
In the local approximation, {\it i.e.} in zero order, we have for the zeroth  
$\tau$-derivative of the direct Born self-energy
\begin{eqnarray}
\sigma_{Bd}^{>,<(0)}({\bf R}, {\bf p} , T ) & = &
{{1}\over{(2\pi )^6}} \int d^3 p_2  \int d^3 p_3
g^{>,<(0)} ({\bf R} , {\bf p}_2, T)
g^{<,>(0)} ({\bf R} , {\bf p}_3, T)
\nonumber \\ & &
g^{>,<(0)} ({\bf R}, {\bf p}- {\bf p}_2+  {\bf p}_3, T)
v^2 ( {\bf p}_2 - {\bf p}_3)
\end{eqnarray}

In a choosen order in the  gradient expansion  for the constraint and kinetic 
equations we need the self-energy in the same order.
It is straightforward to prove that the first order terms in the gradient
expansion of eq. (41) for $\tau =0$ cancel.  Thus, as for the Hartree-Fock part 
of the self-energy, the local approximation for the Born self-energy is
consistent with the gradient approximation for the transport equations. When
considering  second order terms in the gradient expansion for the transport
equations,  also second order terms in the expansion of eq. (41) have to be
retained. The contribution of second order gradient terms to the zero order
$\tau$-derivative of the direct Born self-energy  is of the form
\begin{eqnarray}  &&-{{1}\over{8}} {{1}\over{(2\pi )^3}} \int d^3 p_2
{{1}\over{(2\pi )^3}} \int d^3 p_3 \partial ^R _i \partial ^R _j \left(
g^{>,<(0)} ({\bf R} , {\bf p}_2, T) g^{<,>(0)} ({\bf R} , {\bf p}_3, T) \right)
\nonumber \\
&&g^{>,<(0)} ({\bf R}, {\bf p}- {\bf p}_2+  {\bf p}_3, T)
\partial ^p _i \partial ^p _j v^2 ( {\bf p}_2 - {\bf p}_3)
\end{eqnarray}

In the local approximation, from eq. (42) with eqs.  (7) and eq. (22) one has
\begin{eqnarray}
 i\sigma ^{A(0)} _{Bd}({\bf R} ,{\bf p} , T )  =
{{1}\over{4}}
{{1}\over{(2\pi )^6}} \int d^3 p_2  \int d^3 p_3
\left\{ \left[
g^{K(0)} ({\bf R} , {\bf p}_2, T) g^{K(0)} ({\bf R} , {\bf p}_3, T)+1
\right] \right. \nonumber \\
+  \left.
g^{K(0)} ({\bf R} , {\bf p} - {\bf p}_2 +{\bf p}_3, T)
\left[
g^{K(0)} ({\bf R} , {\bf p}_2, T) -  g^{K(0)} ({\bf R} , {\bf p}_3, T)
\right]\right\} v^2 ( {\bf p}_2 - {\bf p}_3) \nonumber \\
={{1}\over{4}}
{{1}\over{(2\pi )^6}} \int d^3 p_2  \int d^3 p_3
\left[
1 + g^{K(0)} ({\bf R} , {\bf p}_2, T) g^{K(0)} ({\bf R} , {\bf p}_3, T)
\right]
v^2 ( {\bf p}_2 - {\bf p}_3)
\end{eqnarray}
In the second step we assumed that $g^{K(0)} ( {\bf p}) = g^{K(0)} ( -{\bf p}) $ 
which is valid {\it e.g.} in the fireball region of a heavy ion collision.
This step is not necessary but we use it for the following estimate. 
%In the same way we obtain
%\begin{eqnarray}
% \sigma ^{K(0)}(\bf R} ,{\bf p} , T )  =
%{{1}\over{4}}
%{{1}\over{(2\pi )^6}} \int d^3 p_2 \int d^3 p_3
%g^{K(0)} ({\bf R} , {\bf p} - {\bf p}_2 +{\bf p}_3, T)
%\nonumber \\
%\left[
%1 + g^{K(0)} ({\bf R} , {\bf p}_2, T) g^{K(0)} ({\bf R} , {\bf p}_3, T)
%\right]
%v^2 ( {\bf p}_2 - {\bf p}_3)
%\end{eqnarray}
Using the  non-interacting Green functions \cite{THI}  
\begin{eqnarray}
 g _0 ^{>(0)} = -  i (1- n({\bf p} ) ) &,& g _0 ^{<(0)} =  i  n({\bf p} )
\end{eqnarray}
where $ n({\bf p} )$ is the occupation number of the state with momentum
${\bf p}$ and $0< n({\bf p} ) <1$
we obtain an estimate  of 
$i\sigma ^{A(0)}$  in eq. (44)
\begin{eqnarray}
 i\sigma _0 ^{A(0)} & = &
{{1}\over{2}}
{{1}\over{(2\pi )^3}} \int d^3 p_2 {{1}\over{(2\pi )^3}} \int d^3 p_3
\left[
n({\bf p}_2)[1-n({\bf p}_3)] + \right. \nonumber \\ & & \left. n({\bf p}_3)
[1-n({\bf p}_2)]
\right]
v^2 ( {\bf p}_2 - {\bf p}_3)
\end{eqnarray}
We note that  the spectral part of the self-energy
in the Born approximation is an integral of a positive function and thus  
positive.

Eqs. (25,26) in the gradient approximation ($\cos({{1}\over{2}} 
\diamondsuit \to 1$) 
imply
\begin{equation}
g ^A (\tau ) = -i e^{- i \epsilon \tau - {{i \sigma ^{A(0)}}\over{2}}\tau ^2 }
\end{equation}
where we approximate  $\ln g ^A (\tau )$
in second order around $\tau =0$. 
Here the spectral function decays by a gaussian in the relative 
time when the spectral part of the self-energy is calculated using
 eq. (46). Then a  characteristic correlation time is
\begin{equation}
{{1}\over{\tau _{{1}\over{e}}}} = \sqrt{{{i\sigma _0 ^{A(0)}}\over{2}}} \, .
\end{equation}
This estimate can be used to fix the parameter  $\Delta $ 
(or $\alpha$ in Appendix C) in the 
calculation
of the collision integrals, eqs. (29,30),  choosing $\Delta  = 
\tau _{{1}\over{e}}$ (or
${{\alpha}} = \sqrt{{{i\sigma _0 ^{A(0)} }\over{2}}}$) where the spectral 
self-energy, $i\sigma _0 ^{A(0)}$, is determined as above using
the initial occupation numbers.

\vskip 1truecm
\section{Numerical results}

We used eq.(33) (k=1 approximation of the collision integrals)
to test the influence of the terms not usually 
retained in the gradient approximation on the time evolution of 
the phase space nucleon density distribution, $n=(1-ig^{K(0)})/2$.
To keep the
numerical problem as simple as possible we restrict ourselves
to a one-dimensional configuration space (the phase space is two
dimensional); physically, this could be the case of two colliding nuclear
matter slabs. For simplicity, a nucleon-nucleon interaction of the shape of an 
attractive  rectangular well
potential with typical depth (40 $MeV$ ) and range 
(2 $fm$) was used. 

We discretize the partial differential equation, eq. (33), and solve the
resulting finite difference equation. We evolve the phase space 
density distribution  in time steps of 0.1 $fm/c$ ; the density 
distribution is given at mesh points between -8 $fm$ and +8 $fm$
in steps of 0.4 $fm$ in space and between -8 $fm^{-1}$ and +8 $fm^{-1}$
in steps of 0.4 $fm^{-1}$ in momentum. 
The initial density  was taken as a superposition of two
gaussians (the widths were 2 $fm$ in position and 1 $fm^{-1}$ 
in momentum) centered at -2 $fm$  and +2 $fm$ in position and at 
+2 $fm^{-1}$ and -2 $fm^{-1}$ in momentum; the mean momenta of the 
gaussians correspond to an energy of the colliding slabs of 80
$MeV/A$. 

The contour plot of the results of various approximations are 
collected in Fig. 1. The initial phase-space density is
presented in panel (a). To test the program
we calculated first the phase-space density  for free motion, 
i.e. when there is no interaction
and only the first order gradient terms are retained from the drift term
in the left hand side of eq. (33) and the collision terms in the right
hand side of eq. (33) are neglected. The results after 10 $fm/c$ 
are presented in panel (b) 
and they are consistent with the classical Hamiltonian flow 
({\it i.e.} with the test particle method).

The next step is to include the mean field (the Hartree-Fock 
self-energy). The results obtained at 10 $fm/c$ in the gradient 
approximation (Vlasov equation) are presented in panel (c) of Fig.1. 
It is seen that there is considerable distortion of the phase 
space distribution. Including the next order gradient correction 
in the Vlasov equation the results
are given in panel (d). We find that the quantum corrections
to the drift term result in a much smoother behaviour of the 
phase-space distribution function, in fact, more similar to the 
free evolution. The results of 
including the collision terms but with the first order gradient 
expansion is given in panel (e). The hole in the center of phase 
space is filled and the distribution is more extended. This 
calculation corresponds to the usual BUU approach. In panel 
(f) both the next order gradient and the collisions terms are 
included. The density distribution becomes still smoother. 

It is seen that higher order gradient corrections result in a smoother phase
space distribution. This might seem contradictory since with higher 
order gradient corrections one moves closer to the evolution 
of a quantum state, the Wigner transform of which is expected to show
more oscillations. However, it has to be remembered that the initial 
distribution does not correspond to a pure quantum state.

In Fig. 2 we present
the corresponding density distribution in space and momentum; 
they are obtained by integrating the
phase-space distribution over momenta, respectively over space. 
One can see that
the spreading in space of the density distribution increases
by including the third order gradient term in the drift part 
(panel (d) in Fig. 1) 
relative to the Vlasov approximation (c) and 
it increases further when the collisions are included in the calculation (e,f).
The momentum distributions are rather stable with respect to 
the gradient terms (d), but the 
collision terms result in a considerable spreading of the 
distribution in momenta (e,f), as 
expected. With a closer look one sees  that in the Vlasov approximation 
the momentum distribution spreads in a self consistent mean-field (c), 
but this is no 
longer the case  when the gradient (quantum) corrections  are 
taken into account (d).

%One can understand this behaviour as being related to 
%the time evolution of a localized one-body Wigner distribution in an external
%potential. If the distribution spreads in space the corresponding wave
%function will be more extended and, therefore, the Wigner distribution 
%will be narrower in momentum.

These calculations include  nonlocal terms simultaneously in the drift and the
collision part
of the Bolzmann-like equation (33). The inclusion of 
 nonlocality terms in the 
scattering integral given by Morawetz et al. \cite{MOR} leads to consistency
with the thermodynamic virial corrections, but we think that  nonlocal
corrections  should be  also included in the drift term for the case of
nonequilibrium systems.

Up to now we have discussed higher order gradient terms, {\it i.e.} quantum 
corrections to the space evolution. It is of interest to look also 
at the off-shell properties of the Green functions, {\it i.e.} at the properties
of the spectral function. A first look is obtained by discussing 
$g^{A(2)}$, {\it i.e.} the width of the spectral function. According 
to eq. (26) it depends on $\sigma ^{A(0)}$ which we have already calculated.
According to eq. (48) it is also proportional to the inverse of the parameter
$\Delta$, connected with memory effects which are discussed below.
The width of the spectral function is shown for the slab-slab collision in 
Fig. 3 as a function of the one-dimensional distance variable for different 
times, in particular for the initial time (panel (a) of Fig. 1) and for 5 and 
10 fm/c for the BUU approximation (panel (e) of Fig. 1). It is seen that
the width increases in the time evolution of the initially 
nonequilibrated momentum configuration. It is largest during the phase of 
maximum overlap of the slabs. In the later phase of the evolution it becomes
more spread out and lower. This is an expected
behaviour as the 
final configuration should  
evolve towards 
hot nuclear matter (the initial
collective kinetic energy transforms in thermal energy) and the final width
has a contribution from thermal effects.
It is to be noted that the width goes to zero automatically outside
of the system because there the spectral function has to go back on shell.

Another quantum effect on the time evolution of density distribution 
is due to the inclusion of higher order terms in $\Delta$ in the
collision integrals. 
This is a memory  effect as it 
involves time derivatives of the self energies. 
To illustrate numerically this quantum effect we used eq. (35) in the simplest
case, namely two interpenetrating nuclear matter systems  in one dimension, 
{\it i.e.} relative to the previous case the space extension is disregarded. 
In such a 
translationally invariant system all the quantities depend 
only on momentum
and time and eq. (35) can be written
for the the time dependent density in momentum space $n$
\begin{equation}
{{\partial n({\bf p};T)}\over{\partial T}}  = 2\Delta({\bf p};T) \left( 1-
{{\Delta({\bf p};T)}\over{4}} {{\partial}\over{\partial T}}\right) [
(1-n({\bf p};T))(-i \sigma ^{<(0)}({\bf p};T)) -
n({\bf p};T)(i \sigma ^{>(0)}({\bf p};T))]
\end{equation} 
where we used the relations $g^{K(0)}=2i(n-1/2)$
and $g^{A(0)}=-1$ and
 the self-energies in the Born approximation are 
\begin{eqnarray}
i \sigma ^{>(0)}({\bf p};T)&=&{{1}\over{(2\pi) ^6}}\int d^3p_2\int d^3 p_3
v^2({\bf p}_2 - {\bf p}_3) [1-n({\bf p}_2)] n({\bf p}_3)[1-n({\bf p}-
{\bf p}_2 +{\bf p}_3)] \nonumber \\
-i \sigma ^{<(0)}({\bf p};T)&=&{{1}\over{(2\pi) ^6}}\int d^3p_2\int d^3 p_3
v^2({\bf p}_2 - {\bf p}_3) n({\bf p}_2)[1- n({\bf p}_3)]n({\bf p}-
{\bf p}_2 +{\bf p}_3)] \nonumber
\end{eqnarray}
and, using eq. (48) 
\begin{equation}
\Delta({\bf p};T)= \sqrt{{2}\over{(i \sigma ^{>(0)}({\bf p};T))+(-i \sigma
^{<(0)}({\bf p};T))}}
\end{equation}
If we neglect the second order term in $\Delta$ we obtain an equation
%\begin{equation}
%{{\partial n({\bf p};T)}\over{\partial T}}  = 2\Delta({\bf p};T)  [
%(1-n({\bf p};T))(-i \sigma ^{<(0)}({\bf p};T)) -
%n({\bf p};T)(i \sigma ^{>(0)}({\bf p};T))]
%\end{equation} 
which has the structure of a Boltzmann-like equation
\begin{equation}
{{\partial n({\bf p};T)}\over{\partial T}}  = I_{gain}({\bf p};T) -I_{loss}({\bf p};T)
\end{equation} 
with the collision integrals of the Uehling-Uhlenbeck form
\begin{eqnarray}
I_{gain}({\bf p};T) &=&\int d^3p_2\int d^3 p_3 s ({\bf p},{\bf p}_2 - {\bf
p}_3;T) [1-n({\bf p})] [1-n({\bf p}_3)] n({\bf p}_2) n({\bf p}- {\bf p}_2 +{\bf
p}_3) \nonumber \\ 
I_{loss}({\bf p};T) &=&\int d^3p_2\int d^3 p_3 s ({\bf p},{\bf p}_2 -{\bf p}_3;T) 
n({\bf p})[1-n({\bf p}_2)] n({\bf p}_3)[1-n({\bf p}- {\bf p}_2 +{\bf p}_3)]
\nonumber
\end{eqnarray}
but with an effective in-medium cross-section $s ({\bf p},{\bf p}_2 - {\bf p}_3;T)$ 
that depends not only on the transferred momentum, but also on the actual
distribution. Thus  the above equation includes nonequilibrium
effects through the in-medium cross-section
\begin{eqnarray}
s = 
{{2 v^2 \sqrt{2}}\over{(2\pi)^3}}
{{\int d^3p_2'\int d^3 p_3' v^2({\bf p}_2' 
- {\bf p}_3') [n({\bf p}_3')(1- n({\bf p}_2'))+n({\bf p}- 
{\bf p}_2' +{\bf p}_3') (n({\bf p}_2')-n({\bf p}_3'))]}}^{-1/2}
 \nonumber
\end{eqnarray}
which is a complicated functional of $n({\bf p};T)$.
Thus nonequilibrium effects affect the collision
integrals, in the sense that the cross section is dependent on the momentum 
distribution.
In general it is therefore not justified  to use a cross-section calculated for equilibrated configurations in
momentum space in a BUU description of a nuclear collision. 

When we retain the second order term in $\Delta$ in eq. (49), using eq.(50),
we obtain 
\begin{equation}
{{\Delta ^2}\over{2}}\left[ 
{{\partial}\over{\partial T}}(-i\sigma ^{<(0)}({\bf p};T))-n {{\partial}\over{\partial T}}
\left({{2}\over{\Delta ^2}}\right)\right] = I_{gain}({\bf p};T) -I_{loss}({\bf p};T)
\end{equation}
In equilibrium the rhs of eq. (49) vanishes and  one obtains
\begin{eqnarray}
-i\sigma ^{<(0)}_{eq}({\bf p};T) &=& n({\bf p};T){{2}\over{\Delta ({\bf
p};T)^2}} \\ 
i\sigma ^{>(0)}_{eq}({\bf p};T) &=& (1-n({\bf p};T)){{2}\over{\Delta ({\bf
p};T)^2}} 
\end{eqnarray}
Let us remark that if we assume that the condition (53) remains true also in
nonequilibrium (this can be justified if the system is not too far from
equilibrium) and use it in the lhs of eq.(52) we obtain exactly eq. (51). 
More generallly we can write 
\begin{eqnarray}
-i\sigma ^{<(0)}({\bf p};T) &=& n({\bf p};T){{2}\over{\Delta ({\bf
p};T)^2}} +C({\bf p};T)
\end{eqnarray} 
where C is a correction term of the form
\begin{eqnarray}
C({\bf p};T)&=&
{{1}\over{(2\pi) ^6}}\int d^3 p_2 \int d^3 p_3
v^2({\bf p}_2 - {\bf p}_3)\left\{ n({\bf p}_2)[1- n({\bf p}_3)]n({\bf p}  -
{\bf p}_2 +{\bf p}_3)] -
\right. \nonumber \\ && \left.
 n({\bf p};T)
[n({\bf p}_3)(1- n({\bf p}_2))+
n({\bf p}- 
{\bf p}_2 +{\bf p}_3) (n({\bf p}_2)-n({\bf p}_3))]\right\}
\end{eqnarray}
and eq.(52) can be rewritten as
\begin{equation}
{{\partial n({\bf p};T)}\over{\partial T}}  = I_{gain}({\bf p};T) -I_{loss}({\bf p};T)
-{{\Delta({\bf p};T) ^2}\over{2}} {{\partial}\over{\partial T}}C({\bf p};T)
\end{equation} 
which differs from eq.(51) only in the last term which can be considered as a
memory term.

We investigated numerically eq.(51) and (57) in a one-dimensional case starting
with an initial distribution of two gaussians. 
In the case 
$n({\bf p})=n({-\bf p})$ the parameter $\Delta$ does not depend on ${\bf p}$,
but it depends on time. 
The results are presented in Fig. 4. 
One can see that using $\Delta$ calculated at the initial time instead 
of a dynamical $\Delta$ in eq. (51), {\it i.e.} without non-equilibrium
effects on the in medium cross-section, the high momentum distribution tails
are overestimated. Further inclusion of the memory term, eq. (57), changes
mainly the distribuiton at small momenta.

%To illustrate numerically this quantum effect we used eq. (35)  
%in the simplest case, namely two interpenetrating nuclear matters in 
%one dimension. In that case the gradient term vanishes and we compare in 
%Fig. 4 the time evolution of an initial momentum distribution taken as a sum 
%of two gaussians according to eq. (33) (approximation of the collision
%integral  in the lowest order)
%and eq. (35), respectively. We can conclude that the effects
%of the terms neglected in eq. (33) might be important for the velocity 
%of relaxation toward equilibrium and  they result into a slowdown of the
%thermalization.

\vskip 1truecm
\section{Conclusions}

A transport theory for  equal-time
quantities related to the spectral and the kinetic part of the one-particle
Green function has been derived. The present description is not 
in principle restricted
to the mean field approximation and 
higher order terms in the gradient expansion
can be systematically retained.

We note that in the mean field approximation the above
derived equations reduce to the full quantum equations of motion for the
Wigner function in a self-consistently determined mean-field. 
We can then 
 study the effect of
higher order terms in the gradient expansion, {\it i.e.} the quantum corrections 
to the classical equations of motion. This could be of interest
for systems which are localized in space, as is the case for heavy ion
collissions, especially in the initial stage.  The condition of validity
for the gradient  approximation, the product  of the characteristic lengths
at which the Green function varies in position and in momentum is much larger
than $1$ \cite{MRO}, is not fulfilled for a nucleus (one is of the order
of the nuclear dimension, $\sim 1$ {\it fm}, and the other of order of Fermi
momentum, $\sim 1$ {\it fm}$^{-1}$).

We also consider the inclusion of terms which go beyond the mean-field
approximation in the self-energy  and make a  detailed discussion
of the self-energy in the self-consistent Born approximation. 
The compatibility of the two hierarchies
(kinetic and constraint) for the $\tau$-derivatives of the spectral
Green function at  zero relative time (equivalent to the 
energy moments of the spectral function
provided they exist) has been proved up to the second order. 
This extends 
the similar result obtained in the mean-field approximation in
\cite{ETH}.

For the kinetic part of the Green function similar hierarchies of equations
have been derived and the possibility of their systematic truncation 
is discussed.
The lowest order truncation is a Boltzmann-like equation (33) for 
the equal-time
kinetic part of the Green function; all the terms in the gradient expansion
are retained and this allows to study, as in the mean field case,
the effect of the terms neglected
in the gradient approximation. The collision term in eq. (33) has the
Uehling-Uhlenbeck form (see eqs. (42) and  (34)), in which the Pauli 
exclusion principle
is automaticaly taken into account. 

Let us mention that it  is possible 
to prove that 
any truncation of the gradient expansion in eq. (33) conserves the total
number of particles if the first order one, {\it i.e.} the usual gradient
approximation, does. Integrating eq. (33) over position and momentum, 
and using integration by parts to transfer all the derivatives  to the
Green functions (all the selfenergies  are supposed to vanish together with
all their derivatives at infinity, and $\Delta$ is considered constant) the
only surviving term in the rhs of  eq. (33) will be $2 \Delta \int d^3 R \int
d ^3 p  (\sigma ^{A(0)} g ^{K(0)} - \sigma ^{K(0)} g ^{A(0)} )$, {\it i.e.}
the term present in the gradient approximation.
In the gradient approximation, the change in the Boltzmann equation in the
next order truncation of the hierarchies has been derived, eq. (35).

The methods develloped in this work were tested in a simple model of 
one-dimensional slab-on-slab collisions. Here we went only to the lowest 
order of the hierarchy, i.e. we remained in the quasi-particle approximation. 
However, we investigated the effect of the next order gradient terms, i.e. 
of true quantum corrections, and found them of considerable importance. 
Thus these corrections could also influence the dynamical evolution of 
real heavy ion collisions.

The next step will be to consider higher order truncations of the hierarchies.
This will result in a dependence of the time evolution of phase-space density
on the higher order moments of the Green function $g^{K}$. As these moments
are a measure of the off-shellness, this could be an alternative way to
include the off-shell motion in heavy ion collision calculations. For
other proposed approaches in the literature  see \cite{CAS, LEU}.

The approach of the present paper can be used to treat a system
of bosons, or a coupled fermion-boson system, as  quantum hadrodynamics \cite{SER}
which is the modern way for describing the nuclear many-body problem
in the framework of effective field theories.

\newpage
\vskip 1truecm
{\bf Appendix A}

{\bf Properties of the differential operators $\sin {{1}\over{2}} \diamondsuit $

and $\cos {{1}\over{2}} \diamondsuit$}

$$ e^{{{i}\over{2}} \diamondsuit } (f \cdot g) = f \star g$$ is a noncommutative, 
associative product defined on the space of complex valued functions 
$f( {\bf R}, {\bf p})$.

It is necessary to verify the associativity only for functions of the form
$f_i = e^{i ( {\bf \alpha}_i {\bf R} + {\bf \beta}_i  {\bf p} )}$   with 
arbitrary  ${\bf \alpha}_i$, ${\bf \beta}_i$ as every function
can be expressed as a linear combination of the above functions ( Fourier
decomposition) with complex coefficients and the product $\star$ is
obviously distributive.  Using
$$ f_i \star f_j =  e^{{{i}\over{2}} ( {\bf \alpha}_i {\bf \beta}_j - 
{\bf \beta}_i {\bf \alpha}_j )} f_i f_j $$
it is a straightforward exercise to prove that
$$f_1 \star (f_2 \star f_3) =  (f_1 \star f_2 ) \star f_3 \, . $$
We have the identities
$$\cos {{1}\over{2}} \diamondsuit (f \cdot g) = {{1}\over{2}}
(f \star g + g \star f)$$
$$\sin {{1}\over{2}} \diamondsuit (f \cdot g) = {{1}\over{2 i}}
(f \star g - g \star f)$$
and the following relation is obvious

\parbox{12cm}{
\begin{eqnarray}
\sin {{1}\over{2}} \diamondsuit (f \cdot f) = 0  \nonumber
\end{eqnarray}
} \hfill \parbox{2cm}{\begin{eqnarray}(A1) \nonumber  \end{eqnarray}} \newline
Using the definitions of the differential operators and the associativity
we obtain
$$\sin {{1}\over{2}} \diamondsuit (f \cdot \cos {{1}\over{2}}
\diamondsuit (g \cdot h) )=\cos {{1}\over{2}} \diamondsuit (g \cdot \sin {{1}\over{2}}
\diamondsuit (f \cdot h) ) +
\cos {{1}\over{2}} \diamondsuit (h \cdot \sin {{1}\over{2}}
\diamondsuit (f \cdot g) )$$
The particular case of the above relation $g=f$ is

\parbox{12cm}{
\begin{eqnarray}
\cos {{1}\over{2}} \diamondsuit (f \cdot \sin {{1}\over{2}}
\diamondsuit (f \cdot h) ) =
\sin {{1}\over{2}} \diamondsuit (f \cdot \cos {{1}\over{2}}
\diamondsuit (f \cdot h) ) \nonumber
\end{eqnarray}
} \hfill \parbox{2cm}{\begin{eqnarray}(A2) \nonumber  \end{eqnarray}} \newline
We have also the identity
$$\cos {{1}\over{2}} \diamondsuit (g \cdot \cos {{1}\over{2}}
\diamondsuit (f \cdot h) ) +
\sin {{1}\over{2}} \diamondsuit (g \cdot \sin {{1}\over{2}}
\diamondsuit (f \cdot h) ) -
\sin {{1}\over{2}} \diamondsuit (f \cdot \sin {{1}\over{2}}
\diamondsuit (g \cdot h) )$$
$$={{1}\over{4}}\left[
 g \star ( f \star h +h \star f) + ( f \star h +h \star f) \star g
-g \star ( f \star h -h \star f)  \right. $$ $$ +\left. (f  \star h -h \star f)
 \star g
+f \star ( g \star h -h \star g) - ( g \star h -h \star g) \star f
\right] $$
$$={{1}\over{4}}\left[
f \star ( g \star h +h \star g) + ( g \star h +h \star g) \star f
\right] $$
\parbox{13.5cm}{
\begin{eqnarray}
&&=\cos {{1}\over{2}} \diamondsuit (f \cdot \cos {{1}\over{2}}
\diamondsuit (g \cdot h) ) \nonumber
\end{eqnarray}
} \hfill \parbox{2cm}{$(A3)$ }

\vskip 1truecm
{\bf Appendix B}

{\bf Compatibility of the constraint and kinetic hierarchies for the Green 
functions}

Using eq. (22), eq. (A1)  for $f= \epsilon $,
%$ \sin {{1}\over{2}} \diamondsuit (\epsilon  \cdot \epsilon   ) = 0$
  $ {{\partial \epsilon }\over{\partial T}}
= {{\partial \sigma_{\delta} }\over{\partial T}}$  and
 $\cos  {{1}\over{2}}\diamondsuit ( f  \cdot g^{A(0)} ) = f  g^{A(0)} = -i f$
it is a straightforward algebraic exercise to prove that the second kinetic
equation (23) is satisfied by the solution (25) of the first constraint
equation.

To show that the solution of the second constraint equation (26) satisfies
the third kinetic equation (23) we take the action of the operator
$i  {{\partial}\over{\partial T}}  -2 i \sin {{1}\over{2}}
\diamondsuit ( \epsilon \cdot   )$ on the eq. (26) and we obtain
\newline
\parbox{12cm}{
\begin{eqnarray}
&&i  {{\partial}\over{\partial T}} g^{A(2)} -2 i \sin {{1}\over{2}}
\diamondsuit ( \epsilon \cdot  g^{A(2)} )
 =
-2 \cos {{1}\over{2}} \diamondsuit ({{\partial \epsilon }\over{\partial T}}
  \cdot  \epsilon ) -i  {{\partial \sigma^{A(0)}}\over{\partial T}}
\nonumber \\ & &
+2 \sin {{1}\over{2}} \diamondsuit ( \epsilon \cdot
  \cos  {{1}\over{2}}\diamondsuit(  \epsilon  \cdot  \epsilon )            )
+
2 i \sin {{1}\over{2}}
\diamondsuit ( \epsilon \cdot  \sigma ^{A(0)} )  \nonumber \\
&& =
-2 \cos {{1}\over{2}} \diamondsuit ({{\partial \epsilon }\over{\partial T}}
  \cdot  \epsilon )  -i  {{\partial \sigma^{A(0)}}\over{\partial T}}
-2 i \sin {{1}\over{2}}
\diamondsuit ( \sigma ^{A(0)} \cdot \epsilon  ) \, ,\nonumber
\end{eqnarray}
} \hfill \parbox{2cm}{\begin{eqnarray}
\nonumber \\ \nonumber \\ \nonumber \\ \nonumber \\
(B1) \nonumber  \end{eqnarray}} \newline
where in the last step we used the  identity (A2) in the Appendix A
\begin{eqnarray}
 \sin {{1}\over{2}} \diamondsuit ( \epsilon \cdot
  \cos  {{1}\over{2}}\diamondsuit(  \epsilon  \cdot  \epsilon )            ) & = &
\cos {{1}\over{2}} \diamondsuit ( \epsilon \cdot
  \sin  {{1}\over{2}}\diamondsuit(  \epsilon  \cdot  \epsilon ) )  = 0  \nonumber
\end{eqnarray}
The third equation (23) can be rewritten using the relation (25)
\newline
\parbox{12cm}{
\begin{eqnarray}
i  {{\partial}\over{\partial T}} g^{A(2)} -2 i \sin {{1}\over{2}}
\diamondsuit ( \epsilon \cdot  g^{A(2)} )
& = &
-2 \cos {{1}\over{2}} \diamondsuit ({{\partial \epsilon }\over{\partial T}}
  \cdot  \epsilon )  -i  {{\partial \sigma^{A(0)}}\over{\partial T}}
\nonumber \\ & &
-2 i \sin {{1}\over{2}}
\diamondsuit ( \sigma ^{A(0)} \cdot \epsilon  ) \nonumber
\end{eqnarray}
} \hfill \parbox{2cm}{\begin{eqnarray}(B2) \nonumber  \end{eqnarray}} 
\newline 
Comparing eqs. (B1) and (B2) we conclude that the solution (26) of the
constraint hierarchy satisfies the third equation of the kinetic hierarchy.

In the same way we will show the compatibility of the constraint and kinetic
hierarchies for the kinetic part of the Green function in the mean field
approximation.
The second kinetic equation (27) can be obtained from the first 
kinetic equation and the first constraint equation (28). Acting with the
operator
$i  {{\partial}\over{\partial T}}  -2 i \sin {{1}\over{2}}
\diamondsuit ( \epsilon \cdot   )$ on the first constraint equation we obtain
 ($S^K_{\pm}  = 0$)
\begin{eqnarray}
i  {{\partial}\over{\partial T}} g^{K(1)} -2 i \sin {{1}\over{2}}
\diamondsuit ( \epsilon \cdot  g^{K(1)} ) +i (
i  {{\partial}\over{\partial T}}  -2 i \sin {{1}\over{2}}
\diamondsuit  (\epsilon \cdot ) )
\cos {{1}\over{2}} \diamondsuit ( \epsilon \cdot g^{K(0)}  ) & = & 0 \nonumber
\end{eqnarray}
Using the first kinetic equation we obtain
\begin{eqnarray}
i  {{\partial}\over{\partial T}} g^{K(1)} -2 i \sin {{1}\over{2}}
\diamondsuit (\epsilon  \cdot  g^{K(1)} ) -\cos  {{1}\over{2}}
\diamondsuit (  {{\partial \epsilon }\over{\partial T}} \cdot  g^{K(0)} ) -
\nonumber \\
2 \cos {{1}\over{2}} \diamondsuit (\epsilon \cdot \sin {{1}\over{2}}
\diamondsuit (\epsilon \cdot g^{K(0)}) ) +
2 \sin {{1}\over{2}} \diamondsuit (\epsilon \cdot \cos {{1}\over{2}}
\diamondsuit (\epsilon \cdot g^{K(0)}) ) & = & 0 \nonumber
\end{eqnarray}
The last two terms cancel and we obtain exactly the second kinetic
equation (27).

In the same way one can obtain the third kinetic equation. Acting with 
the operator $i  {{\partial}\over{\partial T}}  -2 i \sin {{1}\over{2}}
\diamondsuit ( \epsilon \cdot   )$ on the second constraint equation (28) and
using the first two kinetic equations we obtain
\begin{eqnarray}
i  {{\partial}\over{\partial T}} g^{K(2)} -
2 i \sin {{1}\over{2}}
\diamondsuit (\epsilon  \cdot  g^{K(2)} ) -
\cos  {{1}\over{2}}
\diamondsuit (  {{\partial \epsilon }\over{\partial T}} \cdot  g^{K(1)} ) -
\nonumber \\
{{i}\over{2}} \sin  {{1}\over{2}}
\diamondsuit (  {{{\partial}^2 \epsilon }\over{\partial T ^2}} \cdot  g^{K(0)} ) -
2 \cos {{1}\over{2}} \diamondsuit (\epsilon \cdot \sin {{1}\over{2}}
\diamondsuit (\epsilon \cdot g^{K(1)}) ) +
\nonumber \\
i \cos {{1}\over{2}} \diamondsuit (\epsilon \cdot \cos {{1}\over{2}}
\diamondsuit ( {{\partial \epsilon }\over{\partial T}} \cdot g^{K(0)}) ) +
2 \sin {{1}\over{2}} \diamondsuit (\epsilon \cdot \cos {{1}\over{2}}
\diamondsuit (\epsilon \cdot g^{K(1)}) )+
\nonumber \\
i \sin {{1}\over{2}} \diamondsuit (\epsilon \cdot \sin {{1}\over{2}}
\diamondsuit (  {{\partial \epsilon }\over{\partial T}}   \cdot g^{K(0)}) ) -
i \sin {{1}\over{2}} \diamondsuit ( {{\partial \epsilon }\over{\partial T}} \cdot \sin {{1}\over{2}}
\diamondsuit (\epsilon \cdot g^{K(0)}) )
& = & 0 \nonumber
\end{eqnarray}
The $5^{th}$ and the $7^{th}$ cancel and we obtain
\begin{eqnarray}
i  {{\partial}\over{\partial T}} g^{K(2)} -
2 i \sin {{1}\over{2}}
\diamondsuit (\epsilon  \cdot  g^{K(2)} ) -
{{i}\over{2}} \sin  {{1}\over{2}}
\diamondsuit (  {{{\partial}^2 \epsilon }\over{\partial T ^2}} \cdot  g^{K(0)} ) -
\cos  {{1}\over{2}}\diamondsuit (  {{\partial \epsilon }\over{\partial T}}
 \cdot  g^{K(1)} ) & = &  \nonumber \\
-i \left\{
\cos {{1}\over{2}} \diamondsuit (\epsilon \cdot \cos {{1}\over{2}}
\diamondsuit ( {{\partial \epsilon }\over{\partial T}} \cdot g^{K(0)}) ) +
 \sin {{1}\over{2}} \diamondsuit (\epsilon \cdot \sin {{1}\over{2}}
\diamondsuit (  {{\partial \epsilon }\over{\partial T}}   \cdot g^{K(0)}) ) -
\right. \nonumber \\ \left.
 \sin {{1}\over{2}} \diamondsuit ( {{\partial \epsilon }\over{\partial T}} \cdot \sin {{1}\over{2}}
\diamondsuit (\epsilon \cdot g^{K(0)}) ) \right\}
  =   -i \cos {{1}\over{2}} \diamondsuit ( {{\partial \epsilon }\over{\partial T}} \cdot 
\cos{{1}\over{2}}
\diamondsuit (\epsilon \cdot g^{K(0)}) ) \nonumber
\end{eqnarray}
where we used the  identity (A3) in Appendix A.
Using the first constraint equation (28) we obtain
$$
i  {{\partial}\over{\partial T}} g^{K(2)} -2 i \sin {{1}\over{2}}
\diamondsuit ( \epsilon \cdot  g^{K(2)} ) -
{{i}\over{2}} \sin  {{1}\over{2}}
\diamondsuit (  {{{\partial}^2 \epsilon }\over{\partial T ^2}} \cdot  g^{K(0)} )-
 \cos  {{1}\over{2}}
\diamondsuit (  {{\partial \epsilon }\over{\partial T}} \cdot  g^{K(1)} )
  =  $$
$$ \cos  {{1}\over{2}}
\diamondsuit (  {{\partial \epsilon }\over{\partial T}} \cdot  g^{K(1)} ) $$
which coincides  with the third kinetic equation (27).

\vskip 1truecm
{\bf Appendix C}

{\bf Approximation of integrals appearing in the hierarchies for $g^{K}$}

In the hierarchies of equations for  $g^{K}$ the terms due to collisions,
$S^{K(k)}_{\pm} $ are of the form $I= \int_{-\infty}^0 dx f(x)$.
We will approximate such an integral introducing an arbitrary real and positive
parameter $\alpha$
$$ I = \int_{-\infty}^0 dx \left[ e^{-\alpha x}f(x)\right] e^{\alpha x} \, .$$
If we assume that $e^{-\alpha x}f(x)$ is a function localized around zero (this is 
reasonable
for the functions discussed here depending on the relative time) and $\alpha$ is 
choosen such
that the radius of convergence of the Taylor series for $f$ around zero is greater 
than ${{1}\over{\alpha}}$ ($\alpha >{{1}\over{\rho}}$) then the main
contribution to the integral comes from the region in which the function is
well approximated by its Taylor series in the origin and we have
$$ I = \sum_{n=0}^{\infty} {{1}\over{n!}}{{ {\partial}^n }\over {\partial x^n}}
(  e^{-\alpha x}f(x) ) \mid _{x=0}  \int_{-\infty}^0 dx x^n
e^{\alpha x}$$
Using the notation $f^{(k)} = {{ {\partial}^k }\over {\partial x^k}} f(x)
\mid _{x=0}$  we obtain
\begin{eqnarray}
I= {{1}\over{\alpha}} \sum_{n=0}^{\infty} \sum_{k=0}^{n}
{{n!}\over{k! (n-k)!}} {{f^{(k)}}\over{(-\alpha)^k}} \nonumber
\end{eqnarray}
We consider truncations of the sum over $n$,  $\sum_{n=0}^{N}$, and we will use
$N=1$ or $N=2$ to  approximate  the integral by the derivatives  of $f$
 up to order $N$.

We note that if the radius of convergence of the Taylor series is infinity then
the sum of the above series gives the exact result which does not depend
on $\alpha$. However,  the result of the truncation of the series
depends on the  value of the parameter $\alpha$; a low order approximation
is good if  ${{1}\over{\alpha}}$  is choosen as a characteristic decay time of $f$, 
physically a correlation
time of the system. For example, if we calculate in the $N=2$ approximation
the integral with
$ f=  e^{-\beta ^2 x ^2}$
using $\alpha =\beta$ the approximation will be  ${{1}\over{\beta}}$ which differs
from the exact value ${{\sqrt{\pi}}\over{2}}{{1}\over{\beta}}$ by $10$ percent.

We can use the above results to evaluate  the integrals 
$S^{K(0)}_{\pm} $
and $ S^{K(1)}_{\pm}$. It is
straightforward but tedious   to write the collision terms in order
$N=2$. The results are
\begin{eqnarray}
S^{K(0)}_{\pm}  & = & \pm
  {{3}\over{\alpha}}
\big(1-{{1}\over{2 \alpha}}{{\partial}\over{\partial T}}\big)
( \sigma ^{A(0)} g ^{K(0)} - \sigma ^{K(0)} g ^{A(0)}) \nonumber \\
&&+  {{3}\over{\alpha ^2}}
(\sigma^{A(1)} g^{K(0)}-\sigma^{A(0)} g^{K(1)}
-\sigma^{K(1)} g^{A(0)}+\sigma^{K(0)} g^{A(1)})
   \nonumber
\end{eqnarray}
and
\begin{eqnarray}
& S^{K(1)}_{\pm}   = &
 \frac{1}{2}
(\sigma^{A(0)} g^{K(0)}+\sigma^{K(0)} g^{A(0)} )\nonumber\\
&&
\pm    {{ 1 }\over{3 \alpha}}
\bigg[
\big(
\sigma ^{A(1)} g ^{K(0)} + \sigma ^{A(0)} g ^{K(1)}
-\sigma ^{K(1)} g ^{A(0)} - \sigma ^{K(0)} g ^{A(1)}
\big)
 \nonumber\\
&&
\pm \frac{1}{2}
\bigg(
{{\partial \sigma ^{A(0)} }\over{\partial T}}   g ^{K(0)}
- \sigma ^{A(0)}  {{\partial g ^{K(0)} }\over{\partial T}}
- {{\partial \sigma ^{K(0)} }\over{\partial T}}   g ^{A(0)}
+ \sigma ^{K(0)}  {{\partial g ^{A(0)} }\over{\partial T}}
\bigg) \bigg] 
\nonumber
%\\ &&
%-  \frac{ 3}{2 \alpha ^2}   \bigg[
%\pm \frac{1}{4}
%\bigg(
%{{\partial ^2 \sigma ^{A(0)} }\over{\partial T ^2}} g ^{K(0)}
%- \sigma ^{A(0)} {{\partial ^2 g ^{K(0)} }\over{\partial T ^2}}
%-{{\partial ^2 \sigma ^{K(0)} }\over{\partial T ^2}} g ^{A(0)}
%+ \sigma ^{K(0)} {{\partial ^2 g ^{A(0)} }\over{\partial T ^2}}
%\bigg)   \nonumber\\ &&
%+\bigg(
%{{\partial \sigma ^{A(0)} }\over{\partial T }} g ^{K(1)}
%+ \sigma ^{A(1)} {{\partial  g ^{K(0)} }\over{\partial T }}
%-{{\partial  \sigma ^{K(0)} }\over{\partial T }} g ^{A(1)}
%- \sigma ^{K(1)} {{\partial  g ^{A(0)} }\over{\partial T }}
%\bigg) \nonumber\\
%&& \mp \big(
% \sigma ^{A(2)}  g ^{K(0)} -\sigma ^{A(0)}  g ^{K(2)}
%-\sigma ^{K(2)}  g ^{A(0)} +\sigma ^{K(0)}  g ^{A(2)}
%\big)
%\bigg]  \nonumber
\end{eqnarray}

The similarity of these expansions with the  $\Delta $-expansion used in Chapter 4 
is obvious. It corresponds to replacing ${{1}\over{\alpha}}$ by $\Delta$
with slightly different numerical factors. The reason is also clear: in the
$\Delta $-expansion  we cut
 the integral over the past history
with a sharp cut-off $\Delta $, in the present method by a smooth exponential
cut-off with a parameter $\alpha$.

\vskip 1truecm
{\bf Acknowledgements:} We would like to thank in particular Dr. Rolf Fauser,
whose discussions in the early part of this work were very important.
One of the authors (R. A. I.) thanks
Deutscher Akademischer Austauschdienst (DAAD)  for a fellowship.

%\end{document}

\begin{figure}[p]
\begin{center}
%\epsffile{con_b.eps}
\unitlength1cm
\begin{picture}(14,15)
%\put(0,0){\framebox(13.4,13.5){}}
\put(0,4.5){\makebox{\epsfig{file=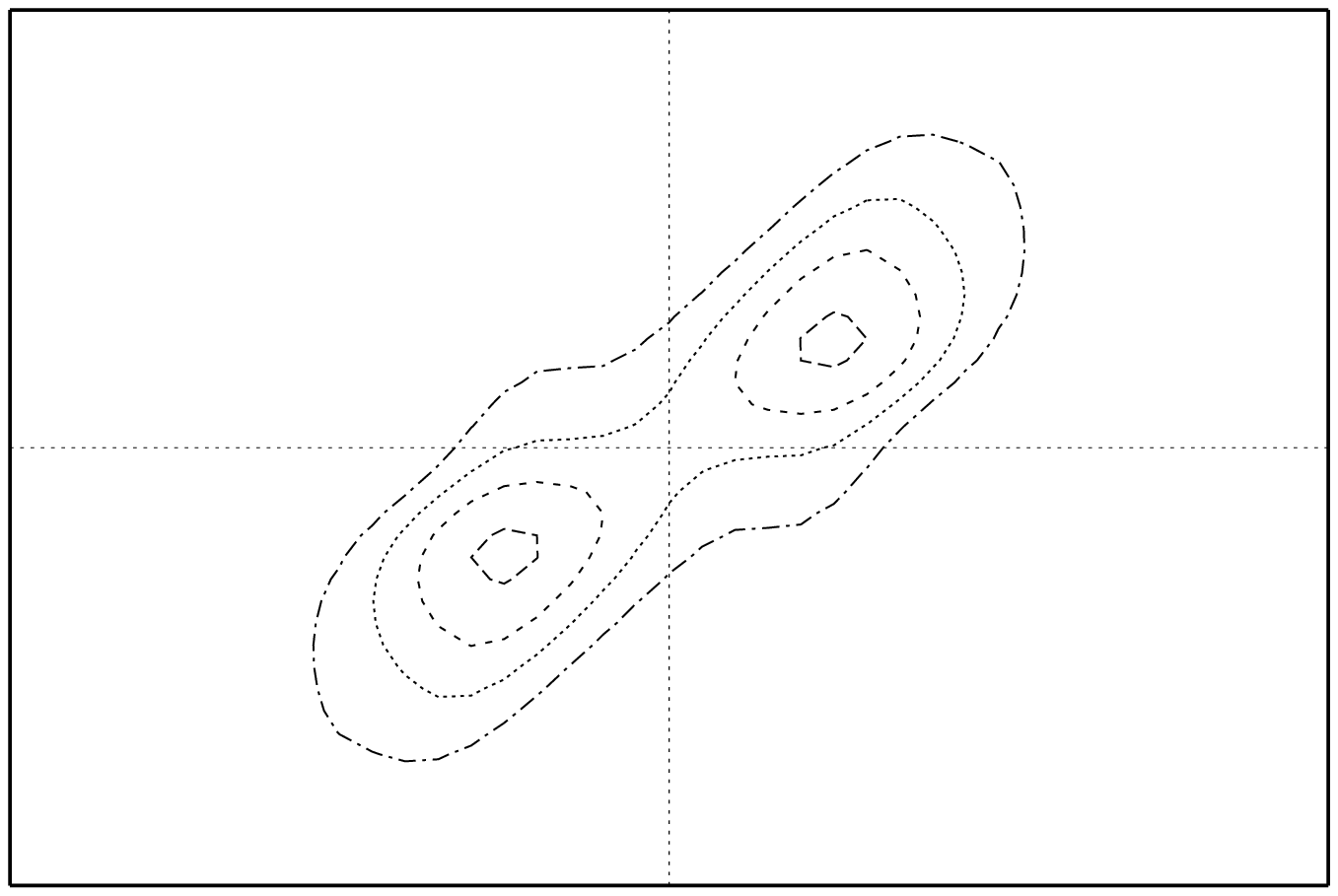,width=4.5cm,angle=-90}}}
\put(6.7,4.5){\makebox{\epsfig{file=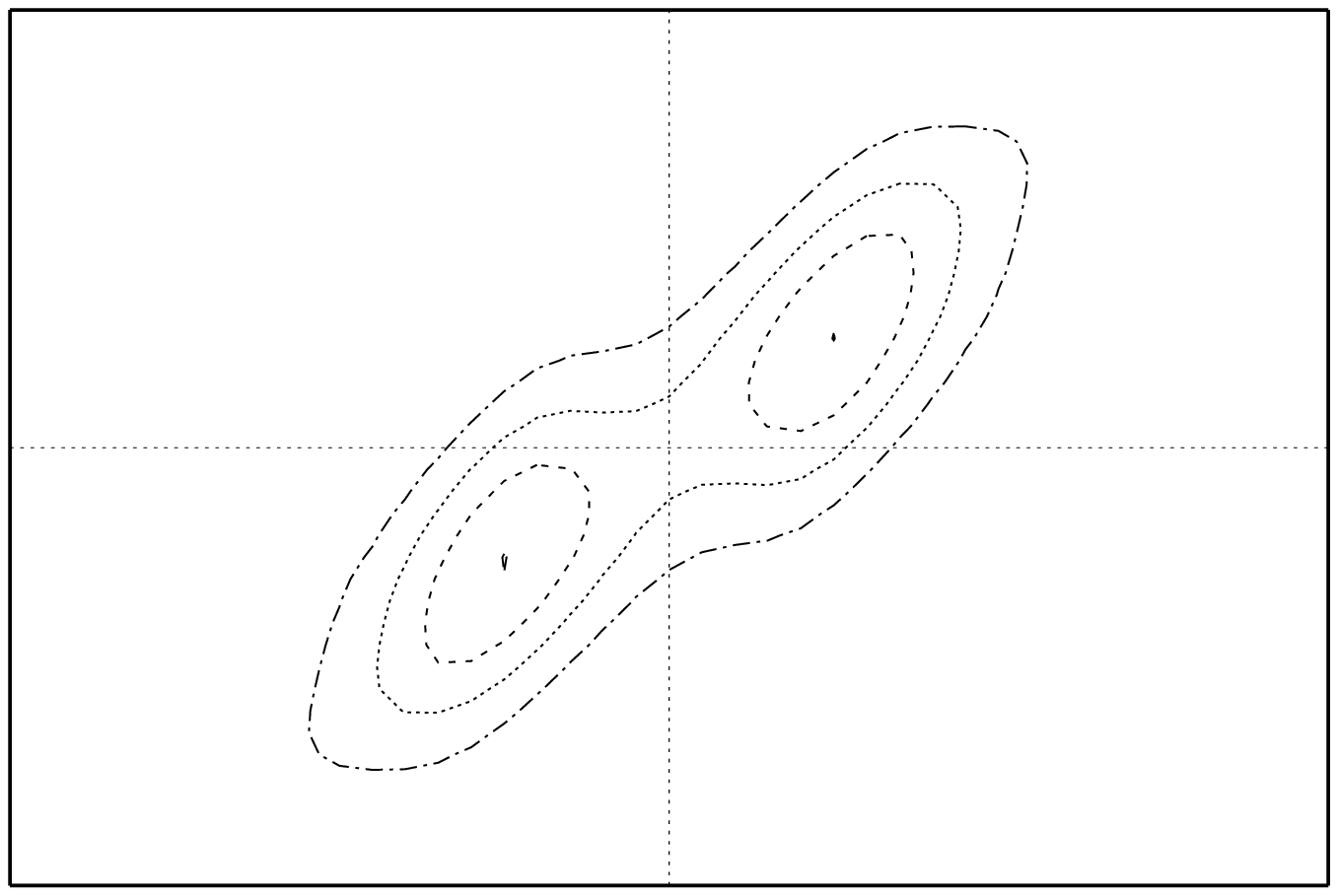,width=4.5cm,angle=-90}}}
%%%%
\put(0,9){\makebox{\epsfig{file=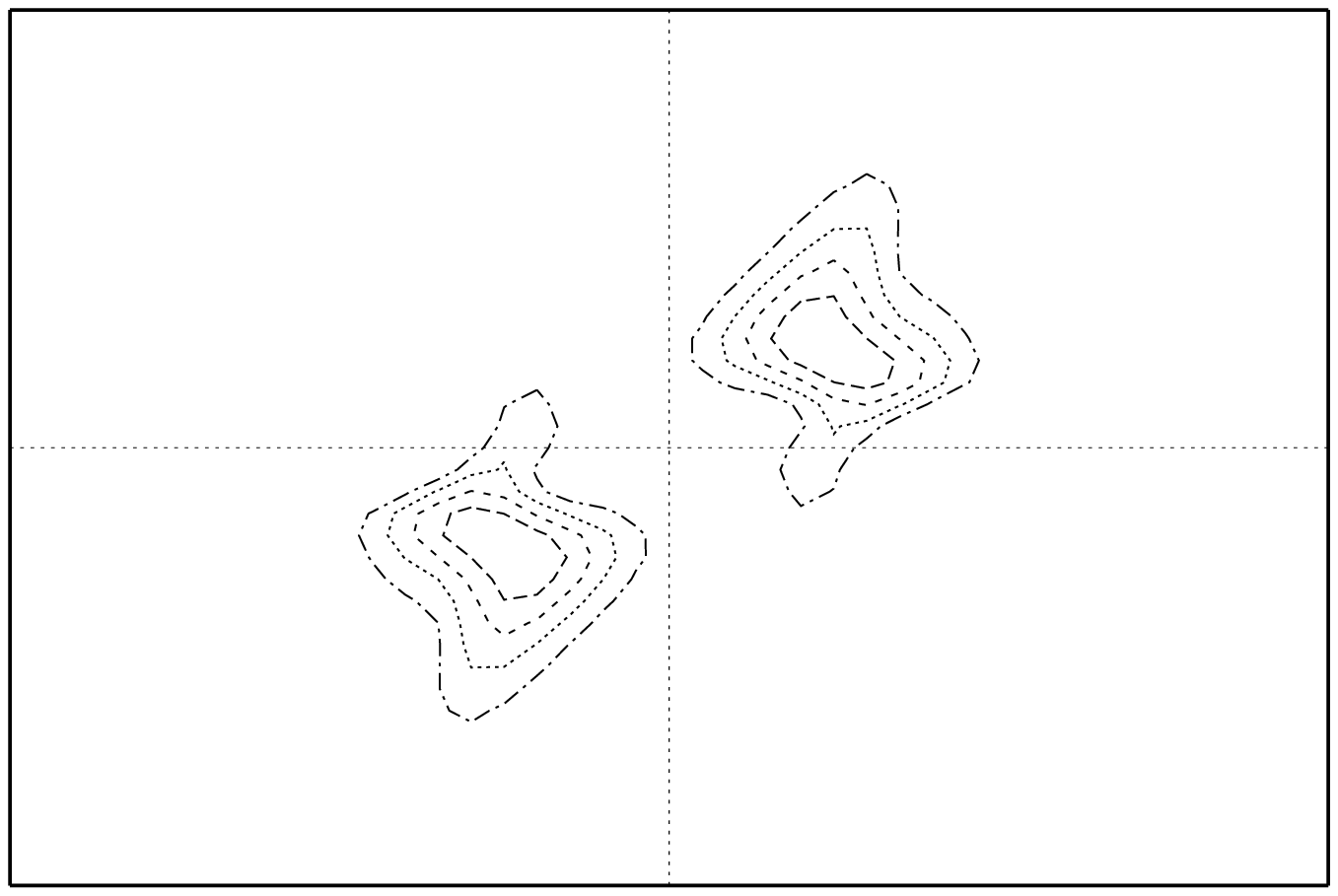,width=4.5cm,angle=-90}}}
\put(6.7,9){\makebox{\epsfig{file=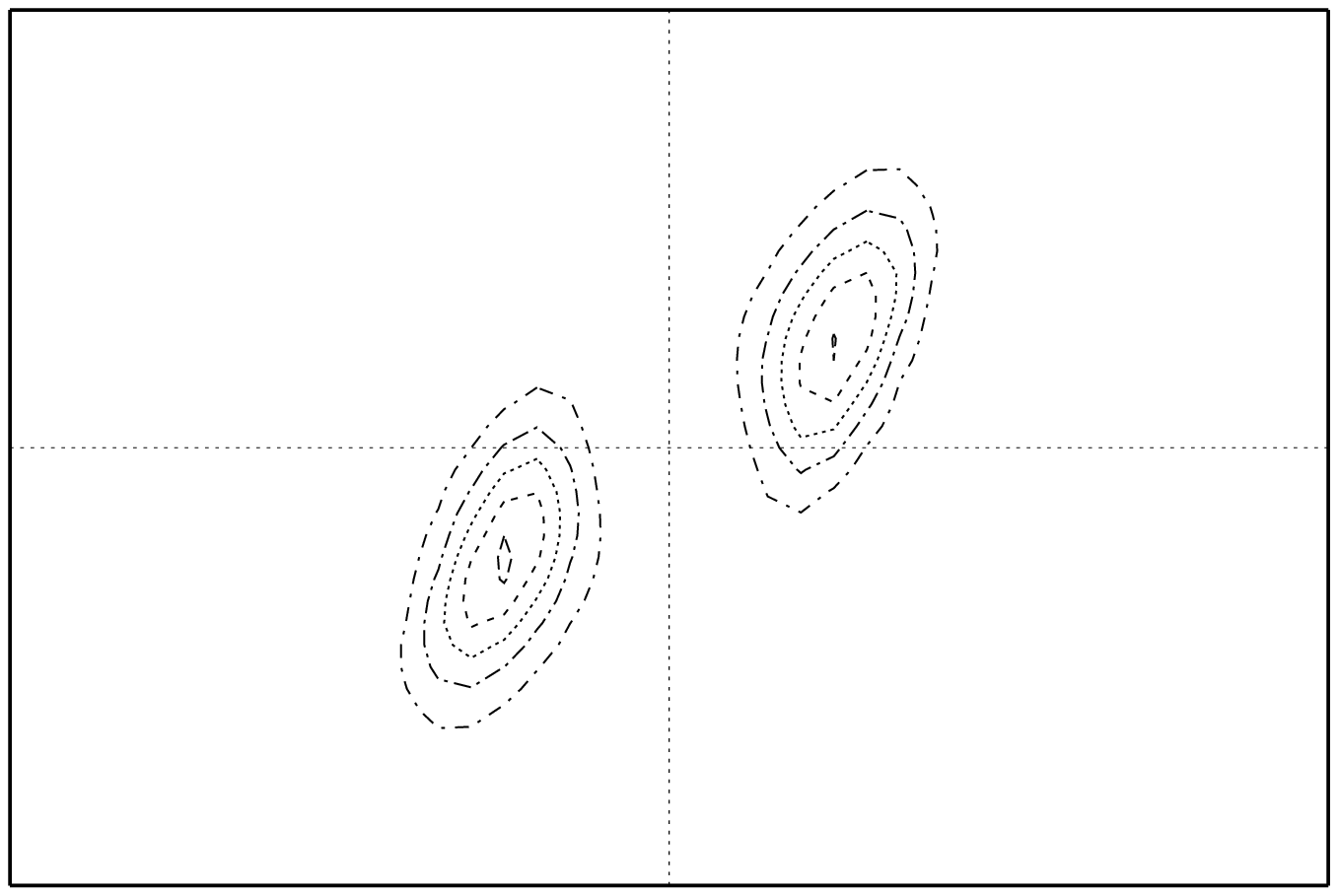,width=4.5cm,angle=-90}}}
%%%%
\put(0,13.5){\makebox{\epsfig{file=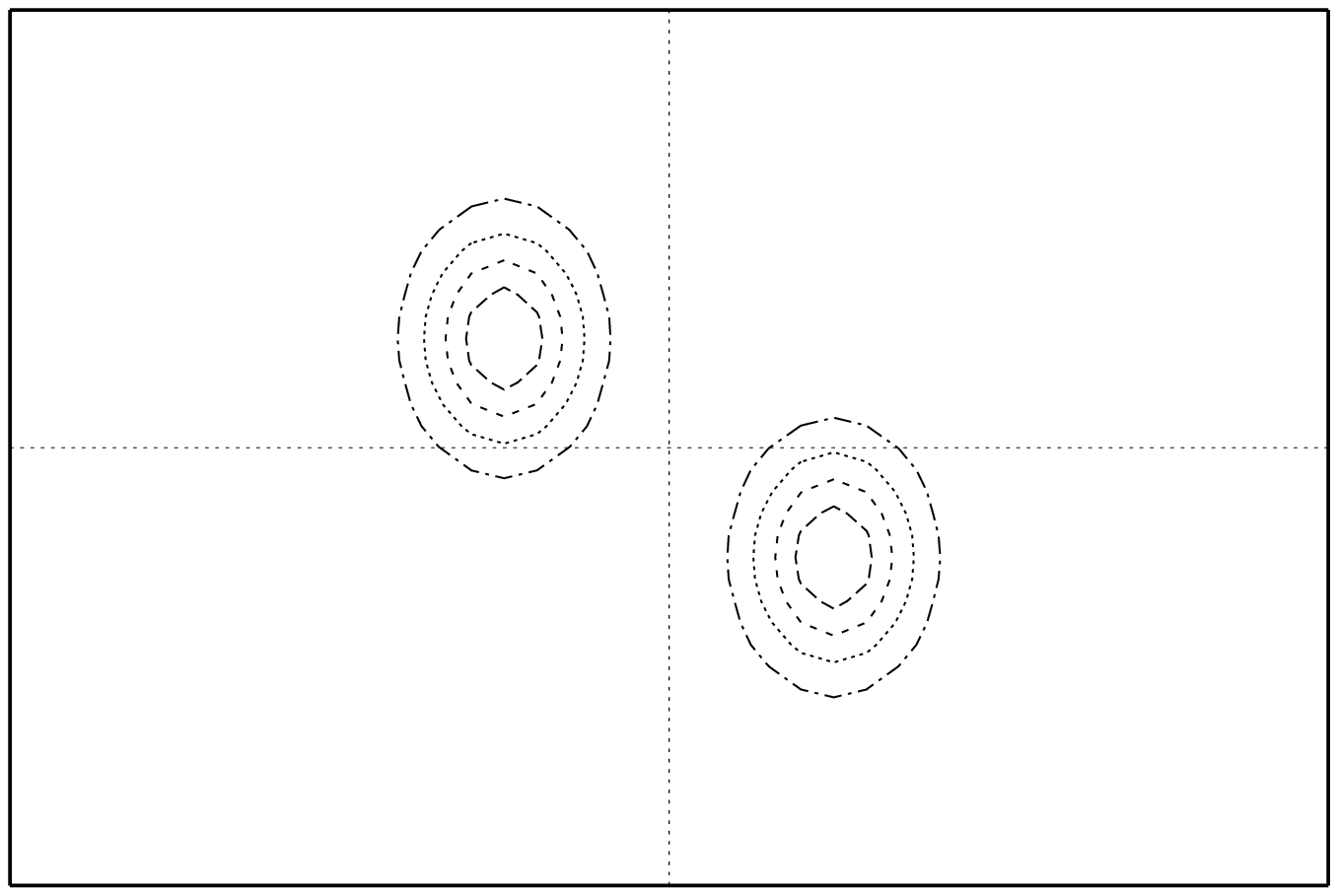,width=4.5cm,angle=-90}}}
\put(6.7,13.5){\makebox{\epsfig{file=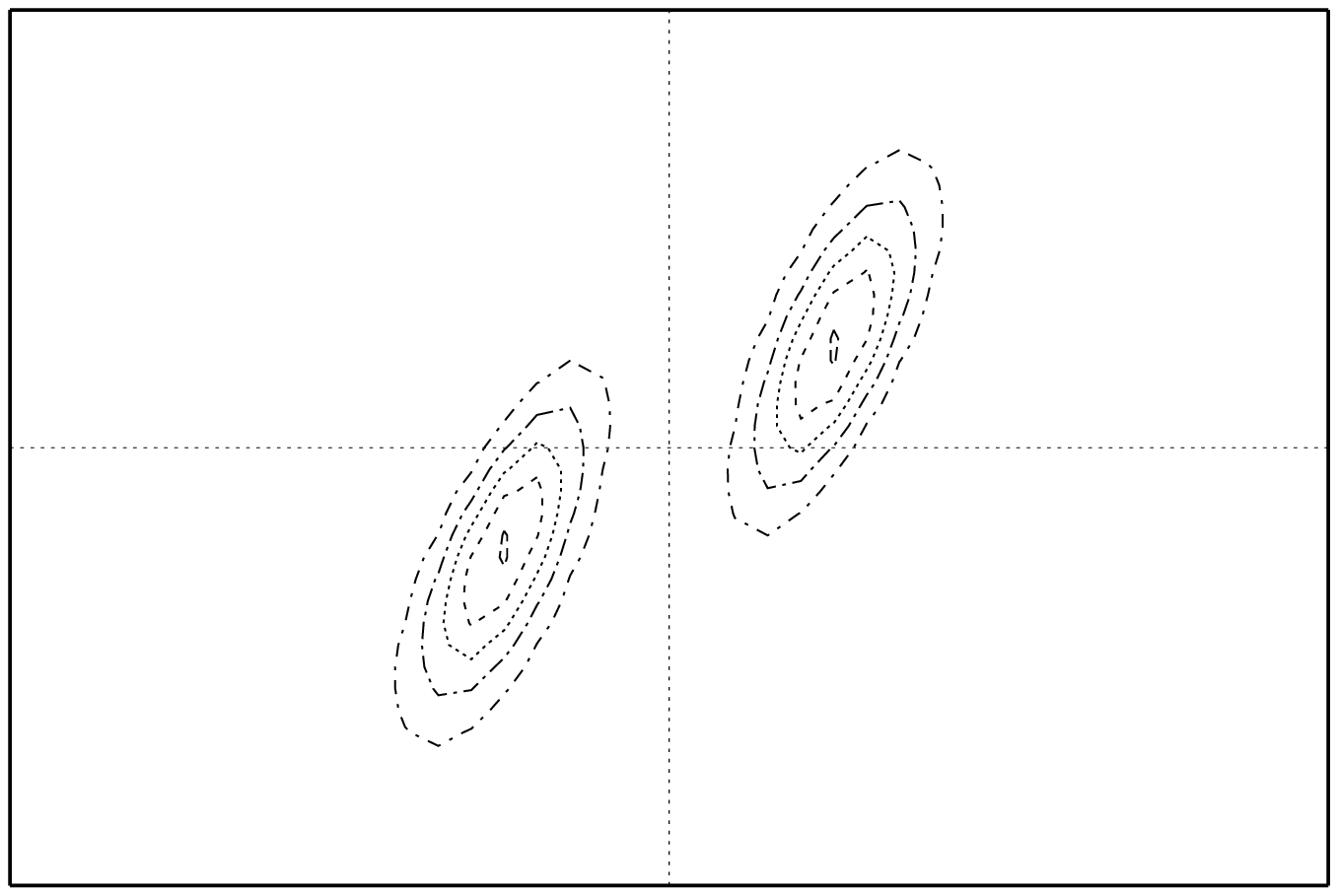,width=4.5cm,angle=-90}}}
\put(5,12.4){\Large a}
\put(11.9,10.2){\Large b}
\put(5,5.7){\Large c}
\put(11.9,5.7){\Large d}
\put(5,1.2){\Large e}
\put(11.9,1.2){\Large f}
\put(1.67,13.3){$|$}
\put(1.67,13.7){$-4$}
\put(3.35,13.7){$0$}
\put(3.35,14.3){$p[fm^{-1}]$}
\put(5.02,13.3){$|$}
\put(5.02,13.7){$+4$}
\put(8.37,13.3){$|$}
\put(8.37,13.7){$-4$}
\put(10.05,13.7){$0$}
\put(10.05,14.3){$p[fm^{-1}]$}
\put(11.72,13.3){$|$}
\put(11.72,13.7){$+4$}
\put(13.2,1.125){$-$}
\put(13.6,1.125){$-4$}
\put(13.6,2.25){$0$}
\put(14,2.25){$x[fm]$}
\put(13.2,3.375){$-$}
\put(13.6,3.375){$+4$}
\put(13.2,5.625){$-$}
\put(13.6,5.625){$-4$}
\put(13.6,6.75){$0$}
\put(14,6.75){$x[fm]$}
\put(13.2,7.875){$-$}
\put(13.6,7.875){$+4$}
\put(13.2,10.125){$-$}
\put(13.6,10.125){$-4$}
\put(13.6,11.25){$0$}
\put(14,11.25){$x[fm]$}
\put(13.2,12.375){$-$}
\put(13.6,12.375){$+4$}
\end{picture}
\caption{\label{contour} Contour plots of the phase space distribution 
of slab-on-slab collisions  at 80 MeV per nucleon displayed for the 
direction of the finite extension of the slabs (x-direction). 
Panel (a) shows the initial distribution. The other panels show the
distributions at time 10 fm/c in the following approximations: (b) free
motion; (c) Vlasov approximation, i.e. first order gradient expansion; 
(d) Vlasov plus next (third) order gradient terms; (e) Vlasov plus 
collision term (corresponds to the usual BUU approximation); 
(f) Vlasov plus next order gradient plus collision terms. 
The contour lines are give in panels (a - d) for 0.2(0.2) and in panels 
(e,f) for 0.1(0.1) with the notation: outermost contour line (increment).
}
\end{center} 
\end{figure}

\begin{figure}[p]
\begin{center}
\unitlength1cm
\begin{picture}(14,20)
%\put(0,0){\framebox(13.4,18){}}
\put(0.5,20){\makebox{\epsfig{file=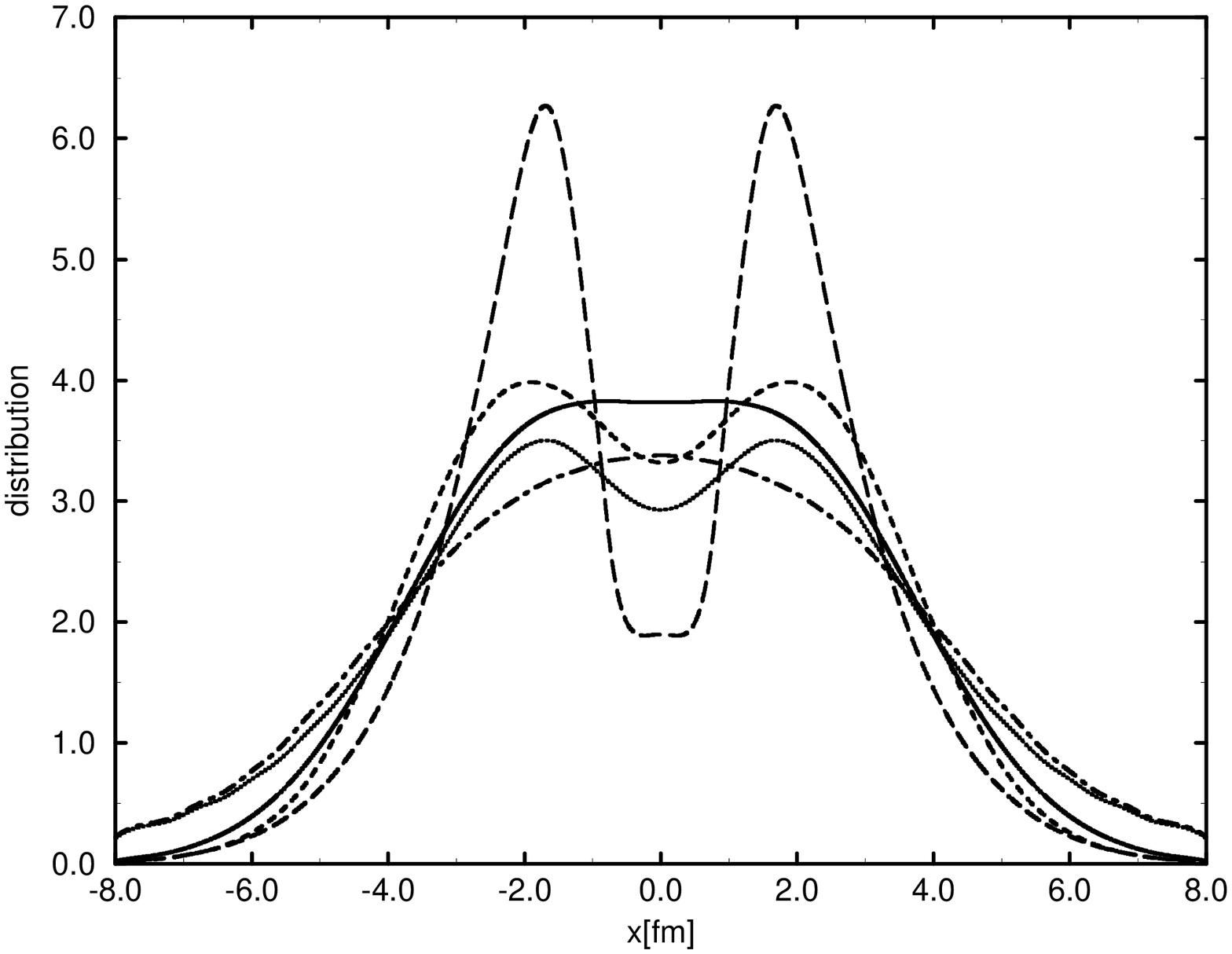,width=10cm,angle=-90}}}
\put(0.5,10){\makebox{\epsfig{file=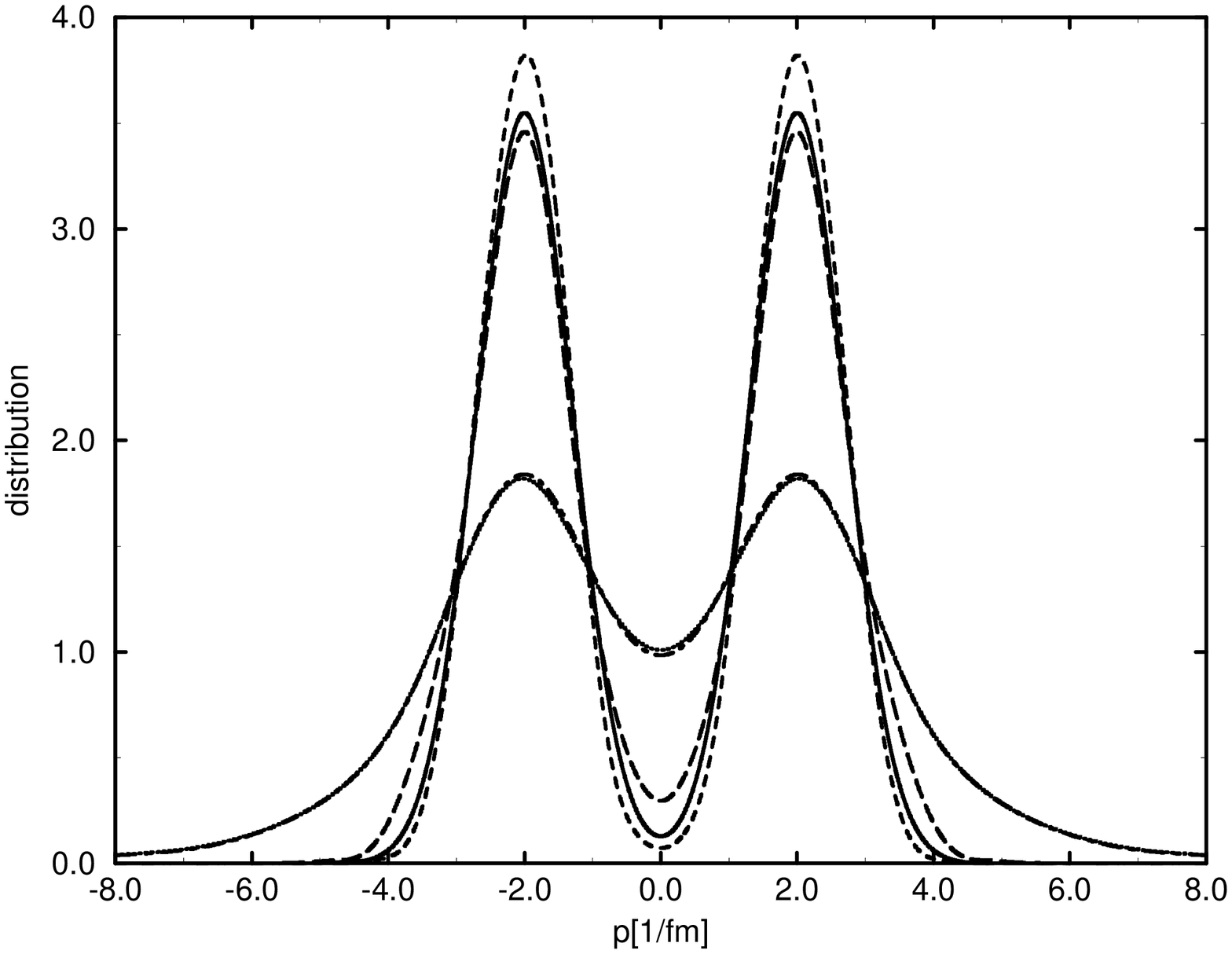,width=10cm,angle=-90}}}
\end{picture}
\caption{\label{distribution} 
Space (top) and momentum (bottom) distributions for the same 
slab-on-slab collisions as in Fig. 1, i.e. the projections on the 
coordinate and momentum axes. The curves correspond to the following
approximations  (with the appropriate panel of Fig.1): free motion 
(solid, panel b); Vlasov (longer dashed, panel c), Vlasov plus next 
order gradient terms (short dashed, panel d), Vlasov plus collision 
terms (dotted, panel e), Vlasov plus next order gradient plus 
collision terms (dash-dotted, panel f). }  
\end{center}
\end{figure}

\newpage
\begin{figure}[p]
\begin{center}
\unitlength1cm
\begin{picture}(14,20)
%\put(0,0){\framebox(13.4,10.5){}}
\put(0.5,10.5){\makebox{\epsfig{file=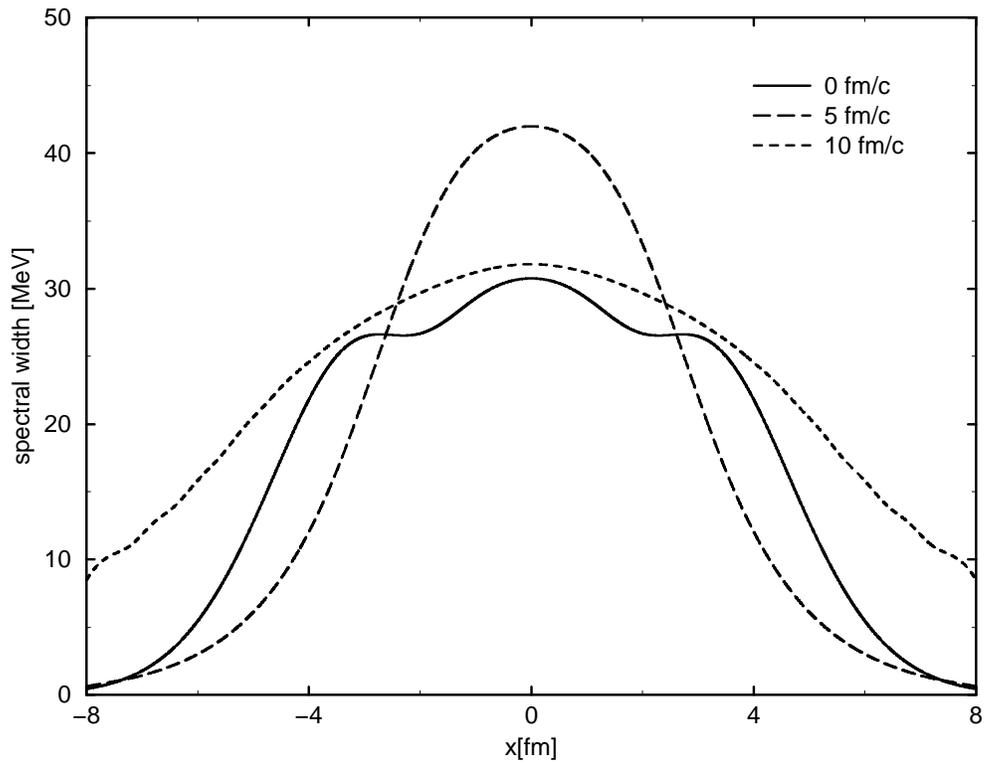,width=10cm,angle=-90}}}
\end{picture}
\caption{\label{width} 
Width (i.e. second order energy moment) of the spectral function at 
different times as a function of the coordinate of the one-dimensional
model. 
} 
\end{center} 
\end{figure}

\newpage
\begin{figure}[p]
\begin{center}
\unitlength1cm
\begin{picture}(14,20)
%\put(0,0){\framebox(13.4,19.5){}}
\put(0.5,20){\makebox{\epsfig{file=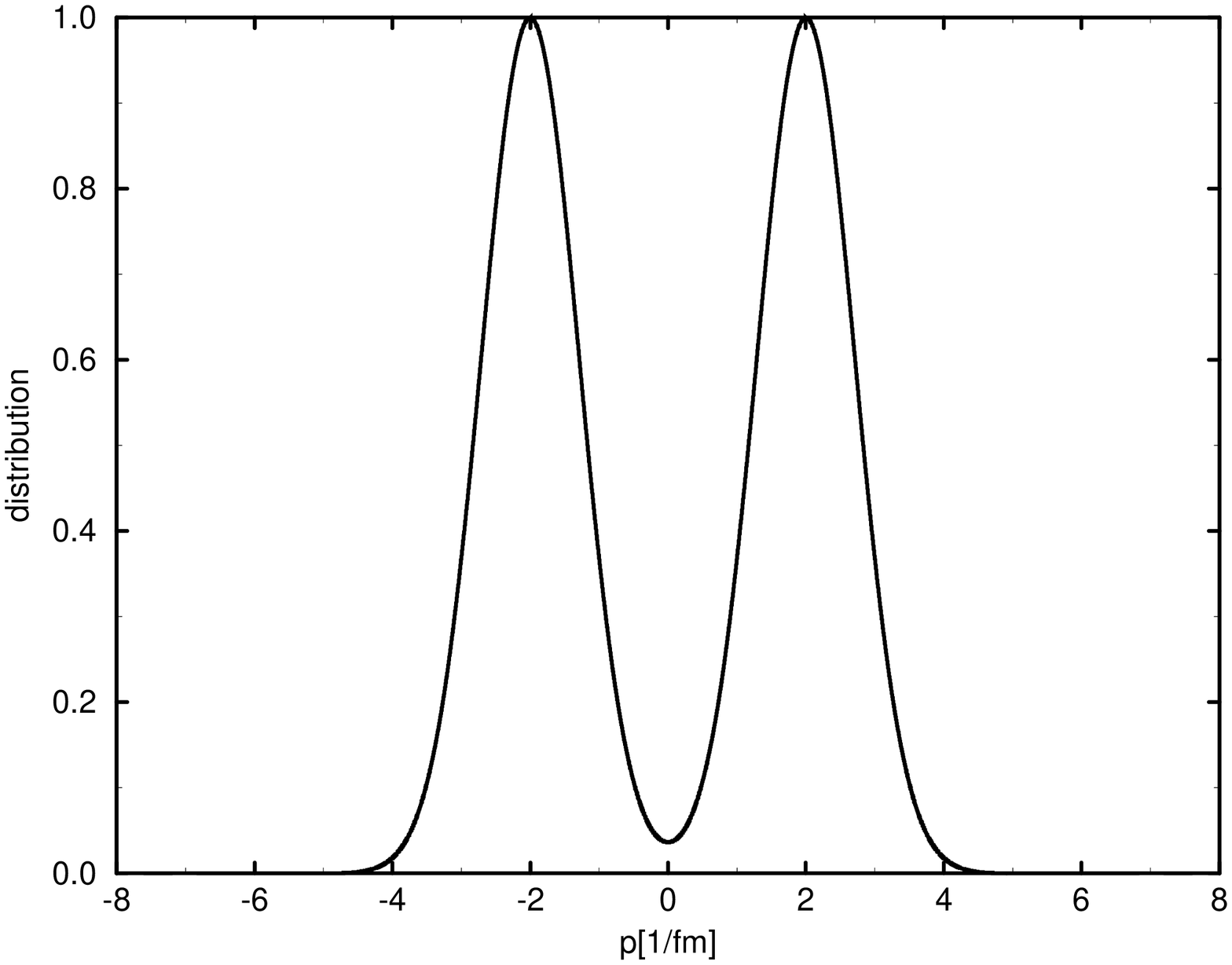,width=10cm,angle=-90}}}
\put(0.5,10){\makebox{\epsfig{file=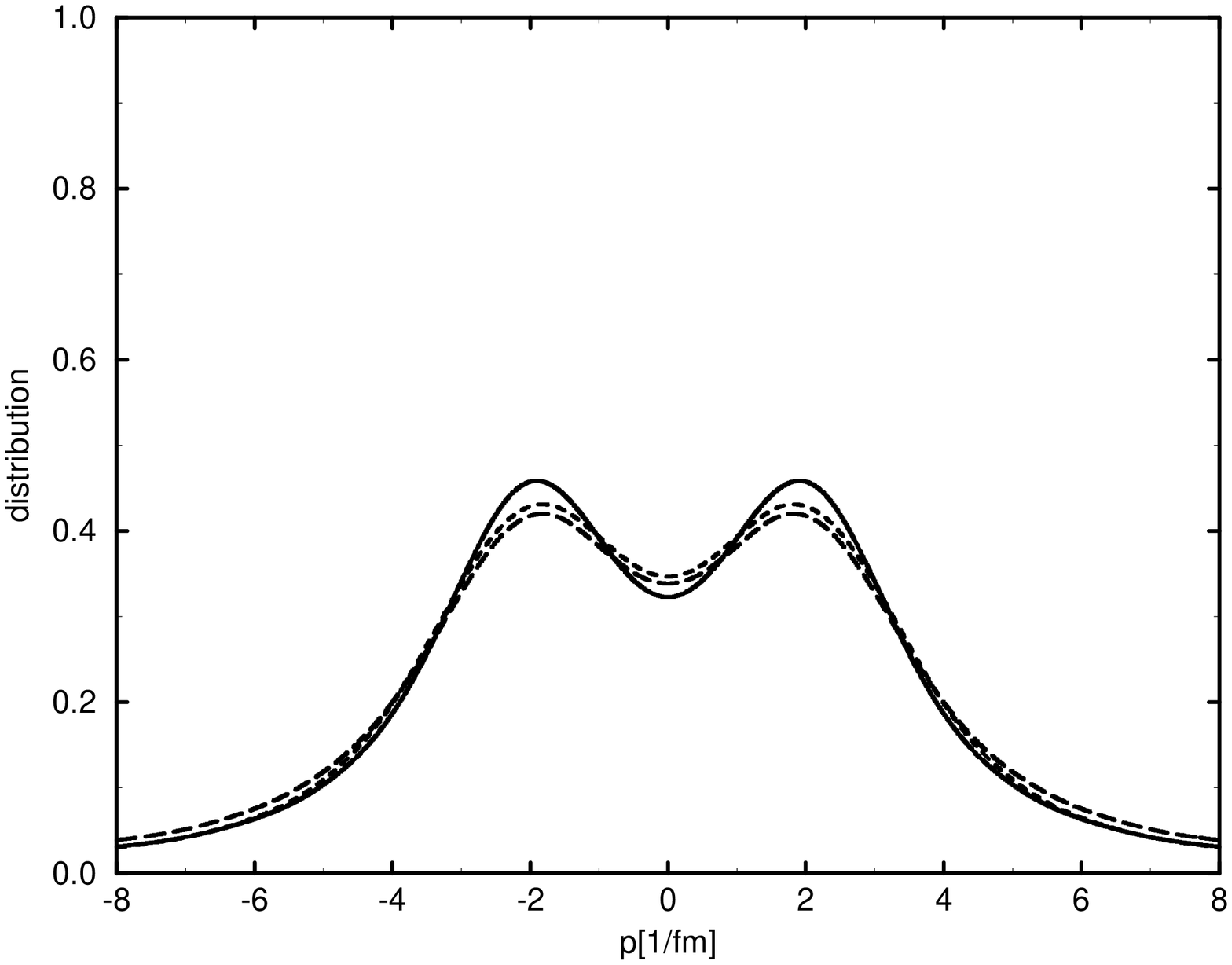,width=10cm,angle=-90}}}
\end{picture}
\caption{\label{thermalization1} 
Time evolution of an initial distribution of two gaussians in momentum (top)
and
at 10 fm/c (bottom). The curves corresponds to the following approximations: 
Boltzmann with time dependent (nonequilibrium) cross-section 
$s ({\bf p},{\bf p}_2 - {\bf p}_3;T)$  (solid
line), Boltzmann with time independent cross-section calculated from the
initial configuration $s ({\bf p},{\bf p}_2 - {\bf p}_3;T=0)$  (long dashed
line), Boltzmann with time dependent cross-section  plus memory term (short
dashed line). 
} 
\end{center} 
\end{figure}

\end{document}